\documentclass[aps,prd,superscriptaddress,showkeys,showpacs,nofootinbib,10pt,longbibliography,twocolumn,floatfix]{revtex4-2}

\usepackage[english]{babel}

\usepackage{color}
\usepackage{graphicx}
\usepackage{amsmath,amssymb,empheq}
\graphicspath{{./Figures/}}
\usepackage{appendix}

\allowdisplaybreaks

\usepackage[breaklinks=true]{hyperref}
\hypersetup{
	colorlinks=true,
	linkcolor=blue,
	filecolor=magenta,      
	urlcolor=blue,
	citecolor=red,
}

\usepackage{orcidlink}

\usepackage{mathrsfs}
\usepackage[normalem]{ulem}

\usepackage{placeins}
\usepackage[utf8]{inputenc}

\input epsf
\usepackage{mathtools}
\usepackage{amsfonts}
\usepackage{upgreek}
\usepackage{latexsym}
\usepackage{stfloats}

\usepackage{afterpage}
\usepackage{bm}

\usepackage{siunitx}
\usepackage{multirow}

\usepackage{stackengine}
\usepackage{scalerel}
\usepackage{xcolor}

\begin{document}

{\hfill MS-TP-25-28}
\title{Vector dark matter production during inflation in the gradient-expansion formalism}

\author{A.V.~Lysenko\,\orcidlink{0009-0001-9205-0122}}
\email{lysenkoktp@knu.ua}
\affiliation{Physics Faculty, \href{https://ror.org/02aaqv166}{Taras Shevchenko National University of Kyiv}, 64/13, Volodymyrska Street, 01601 Kyiv, Ukraine}

\author{O.O.~Sobol\,\orcidlink{0000-0002-6300-3079}}
\email{oleksandr.sobol@knu.ua}
\affiliation{Institute for Theoretical Physics, \href{https://ror.org/00pd74e08}{University of M\"{u}nster}, Wilhelm-Klemm-Stra{\ss}e 9, 48149 M\"{u}nster, Germany}
\affiliation{Physics Faculty, \href{https://ror.org/02aaqv166}{Taras Shevchenko National University of Kyiv}, 64/13, Volodymyrska Street, 01601 Kyiv, Ukraine}
	
\author{S.I.~Vilchinskii\,\orcidlink{0000-0002-9294-9939}}
\email{stanislav.vilchinskii@cern.ch}
\affiliation{D\'{e}partement de Physique Th\'{e}orique and Center for Astroparticle Physics, \href{https://ror.org/01swzsf04}{Universit\'{e} de Gen\`{e}ve},  24 quai Ernest Ansermet, 1211 Gen\`{e}ve 4, Switzerland}
%\affiliation{Niels Bohr Institute, \href{https://ror.org/035b05819}{University of Copenhagen}, Blegdamsvej 17, 2010 Copenhagen, Denmark}
\affiliation{Physics Faculty, \href{https://ror.org/02aaqv166}{Taras Shevchenko National University of Kyiv}, 64/13, Volodymyrska Street, 01601 Kyiv, Ukraine}

\date{\today}

%%%%%%%%%%%%%%%%%%%%%%%%%%%%%%%%%%%%%%%%%%%%%%%%%%%%%%%%%%%%%%%%%%%%%%%%%%
\begin{abstract}
    A   massive vector field is a highly promising candidate for dark matter in the Universe. A salient property of dark matter is its negligible or null coupling to ordinary matter, with the exception of gravitational interaction. This poses a significant challenge in producing the requisite amount of dark particles through processes within the Standard Model. In this study, we examine the production of a vector field during inflation due to its direct interaction with the inflaton field through kinetic and axionlike couplings as well as the field-dependent mass. The gradient-expansion formalism, previously proposed for massless Abelian gauge fields, is extended to include the longitudinal polarization of a massive vector field. We derive a coupled system of equations of motion for a set of bilinear functions of the vector field. This enables us to address the nonlinear dynamics of inflationary vector field production, including backreaction on background evolution. To illustrate this point, we apply our general formalism to a low-mass vector field whose kinetic and mass terms are coupled to the inflaton via the Ratra-type exponential function. The present study investigates the production of its transverse and longitudinal polarization components in a benchmark inflationary model with a quadratic inflaton potential. It has been demonstrated that pure mass coupling is able to enhance only the longitudinal components. By turning on also the kinetic coupling, one can get different scenarios. As the coupling function decreases, the primary contribution to the energy density is derived from the transverse polarizations of the vector field. Conversely, for an increasing coupling function, the longitudinal component becomes increasingly significant and rapidly propels the system into the strong backreaction regime.
\end{abstract}
%%%%%%%%%%%%%%%%%%%%%%%%%%%%%%%%%%%%%%%%%%%%%%%%%%%%%%%%%%%%%%%%%%%%%%%%%%

\maketitle

%%%%%%%%%%%%%%%%%%%%%%%%%%%%%%%%%%%%%%%%%%%%%%%%%%%%%%%%%%%%%%%%%%%%%%%%%%
\section{Introduction}
\label{sec:intro}

According to observational data, dark matter (DM) is the most abundant form of matter in the Universe, accounting for roughly one quarter of its total energy density. The existence of DM is supported by numerous independent astrophysical observations, including galaxy rotation curves, gravitational lensing, the dynamics of hot gas in clusters, matter clustering during large-scale structure formation, primordial nucleosynthesis, and measurements of the cosmic microwave background (CMB). All of these indicate the presence of an invisible, massive component that interacts primarily through gravity~\cite{Bertone:2016nfn,Arbey:2021gdg,Cirelli:2024ssz,Buckley:2017ijx}.
Data from primordial nucleosynthesis and the cosmic microwave background further establish that DM must be nonbaryonic in nature. The Standard Model (SM) of particle physics cannot account for the totality of DM: the only viable SM candidates, neutrinos, are too light (constituting hot DM) and are tightly constrained by structure formation and CMB data to contribute no more than a few percent of the total DM abundance. Thus, the composition of the dominant DM component remains unknown, making its nature a central open problem in both cosmology and particle physics.
A wide range of extensions of the SM propose various DM candidates~\cite{Bertone:2004pz,Asaka:2005an,Roszkowski:2017nbc,Carr:2020xqk,Famaey:2011kh,Fabbrichesi:2020,Adams:2022pbo,Li:2013nal,Belanger:2022esk,Arbey:2020ldf}. Among them, the most widely studied are weakly interacting massive particles, axions, and dark photons, as they are also motivated by considerations beyond DM. Other possibilities include primordial black holes, self-interacting DM, fuzzy DM, asymmetric DM, and $Q$-balls (for a recent comprehensive review, see Ref.~\cite{Cirelli:2024ssz}).

The vector DM, or the massive dark photon~\cite{Fabbrichesi:2020, Caputo:2021eaa},%
\footnote{There are two kinds of dark photons: massless
and massive ones; their theoretical frameworks, experimental and cosmological signatures are distinct. In this paper, we consider only the case of a massive dark photon which is a relevant DM candidate.}
could potentially explain not only the existence of DM, but also dark energy, as well as some other anomalies observed in particle physics experiments~\cite{Hewett:2012ulb,Raggi:2015yfk,Deliyergiyev:2015oxa,Alekhin:2015byh,Curciarello:2016jbz,Alexander:2016aln,Beacham:2019nyx}.
The Lagrangian of this theory resembles that of electromagnetism, but includes both a mass term and a kinetic mixing term with SM photons~\cite{Holdom:1985ag}. It can be viewed as a framework similar to SM electromagnetism featuring one or more additional vector particles that couple to the electromagnetic current. The conservation of this current protects the vector mass term\,---\,or, more generally, the mass matrix of the vector states\,---\,from additive renormalization. If the determinant of the mass matrix vanishes, the theory necessarily contains one massless vector, identified as the photon, along with several massive vectors, commonly referred to as dark photons.
In the minimal dark photon model, the SM is extended by just one additional $U(1)_\mathrm{D}$ gauge group. The corresponding Lagrangian has the form:
\begin{align}
		\label{S1}
		\mathcal{L}\supset -\frac{1}{4}F_{\mu\nu}F^{\mu\nu}-\frac{1}{4}F^\mathrm{D}_{\mu\nu}{F}_\mathrm{D}^{\mu\nu} -\frac{\epsilon}{2}F_{\mu\nu}{F}_\mathrm{D}^{\mu\nu}+
        e A^\mu J_\mu\nonumber\\
        +e_\mathrm{D} A_\mathrm{D}^\mu J^\mathrm{D}_\mu
        +
        \frac{\epsilon\, e}{\sqrt{1-\epsilon^2}} A_\mathrm{D}^\mu J_\mu
        +\frac{m^{2}}{2} A^\mathrm{D}_{\mu}A_\mathrm{D}^{\mu},
\end{align}
where $A_\mu$ and $F_{\mu\nu}$ are the four-potential and field tensor for the SM electromagnetic field, $J_\mu$ is the charged matter current, $e$ is the gauge coupling constant; the same notations with the subscript or superscript ``D'' denote the corresponding quantities in the dark sector; $m_\mathrm{D}$ is the dark photon mass and $\epsilon$ is the kinetic mixing parameter.
In its simplest realization without dark current, the model is fully specified by the dark photon mass and the kinetic mixing parameter (in addition to the SM parameters which are known) which suffice to determine the production cross sections and decay properties of the dark photon. The kinetic mixing acts as a portal between the visible and dark sectors, offering a possible pathway for experimental detection of dark photons. To date, however, no dark photon has been observed, and it remains a hypothetical particle. Nonetheless, experimental searches, together with cosmological and astrophysical data, place significant constraints on its mass and kinetic mixing parameter~\cite{McDermott:2019lch, Caputo:2021eaa,Aramburo-Garcia:2024cbz,Cyncynates:2024yxm}.

A very light dark photon with a small but nonzero mass can serve as a dark matter candidate if it is generated nonthermally in the early Universe. In one class of scenarios~\cite{Arias:2012az,Nelson:2011sf}, the dark photon mass is generated through the St\"{u}ckelberg mechanism, which requires a nonminimal coupling to gravity. When the Hubble parameter decreases below the dark photon mass, the field begins to oscillate, and these oscillations behave as nonrelativistic matter, effectively acting as cold dark matter.
An alternative possibility is that dark photons are produced from vacuum fluctuations during inflation~\cite{Graham:2015rva,Himmetoglu:2008hx,Bastero-Gil:2018uel,Ema:2019yrd,Nakai:2020cfw,Kolb:2020fwh,Ahmed:2020fhc,Salehian:2020asa,Barman:2021qds,Arvanitaki:2021qlj,Sato:2022jya,Bastero-Gil:2022fme,Redi:2022zkt,Barrie:2022mma,Cembranos:2023qph,Cyncynates:2023zwj,Ozsoy:2023gnl,Grzadkowski:2024oaf,Capanelli:2024pzd,Marriott-Best:2025sez,LaRosa:2025woi}. Unlike the electromagnetic field, a massive vector field can be generated without requiring a nonminimal coupling to gravity or the inflaton. This is possible because the presence of a mass term in the Lagrangian automatically breaks conformal invariance of the action\,---\,a property that standard electromagnetism does not possess. However, in such cases, only the longitudinal modes are efficiently generated; see, e.g., Refs.~\cite{Graham:2015rva,Ahmed:2020fhc,Arvanitaki:2021qlj}. To achieve substantial production of transverse modes, one must introduce an interaction either with the spacetime curvature~\cite{Karciauskas:2010as,Kolb:2020fwh,Barman:2021qds,Cembranos:2023qph,Capanelli:2024pzd,Grzadkowski:2024oaf,Ozsoy:2023gnl} or with the inflaton field~\cite{Firouzjahi:2020whk,Nakayama:2020rka,Nakai:2020cfw,Salehian:2020asa,Bastero-Gil:2018uel,Bastero-Gil:2021wsf,Barrie:2022mma,Redi:2022zkt,Cyncynates:2023zwj}, as is commonly done in analogous scenarios of inflationary magnetogenesis (for a review, see Refs.~\cite{Turner:1987bw,Durrer:2013pga,Subramanian:2015lua}).

In this work, we focus on the latter mechanism and consider a single massive Abelian vector field interacting with the inflaton field via generic kinetic (or dilatonic) coupling, axial coupling, as well  through the inflaton-dependent mass. These couplings are included simultaneously in order to cover a large range of scenarios considered in the literature. We are paying special attention to the regime of strong backreaction of inflationary massive vector field production.

At the same time, we neglect the kinetic mixing of the vector DM field with the ordinary electromagnetic field. In this way, we can study the production of the dark vector field fully decoupled from the SM sector. Inclusion of the kinetic mixing is an important next step, which we intend  to address in future work.

The generation of vector dark matter is usually described using the method of mode expansion in Fourier space, where the evolution of each individual mode is studied independently of the others. However, this approach becomes inadequate when the generated fields are too strong, so that they start impacting the inflationary dynamics. In such cases, the strong backreaction regime occurs~\cite{Salehian:2020asa, Bastero-Gil:2022fme}, and the evolution of each mode depends on all the others. The equations thus become nonlinear, requiring a different mathematical framework capable of accounting for these nonlinearities.

A suitable approach is the gradient-expansion formalism (GEF). Instead of working in Fourier space, this method considers a set of quadratic functions of the vector field directly in coordinate space. These functions simultaneously capture all physically relevant modes, allowing the method to remain applicable even in the presence of nonlinear dynamics. The GEF has previously been applied with success to various problems in inflationary magnetogenesis~\cite{Sobol:2020lec, Gorbar:2021rlt, Gorbar:2021ajq, Durrer:2023rhc, vonEckardstein:2023gwk, Gorbar:2023zla, Domcke:2023tnn, vonEckardstein:2024tix, Lysenko:2025pse}. However, it has never been used to describe the generation of massive vector dark matter. One of the aims of this work is to extend the GEF\,---\,originally developed for massless gauge fields\,---\,to incorporate the longitudinal polarization of massive vector fields. 

The remainder of the paper is organized as follows. In Sec.~\ref{sec:model}, we introduce our model of vector dark matter coupled to the inflaton field during inflation. Section~\ref{sec:transverse} is devoted to the evolution of the transverse polarization components of the vector field, analyzed both in Fourier space and in position space; in the latter case, we derive the equations of the GEF and supplement them with the appropriate boundary terms. In Sec.~\ref{sec:longitudinal}, we carry out a similar analysis for the longitudinal polarization component of the vector field. All resulting equations of motion are summarized in Sec.~\ref{sec:system}, where we also discuss the couplings between different sectors and their impact on the inflationary background. In Sec.~\ref{sec:results}, we specify the coupling functions within a simple benchmark scenario and present numerical results for several parameter choices. For the case of negligible backreaction, we validate the GEF against the mode-by-mode solution in Fourier space. We then explore the regime of strong backreaction. Finally, in Sec.~\ref{sec:conclusions}, we provide our conclusions and outline directions for future research. In the Appendix~\ref{app:longitudinal} we discuss in more detail the equations of motion for the longitudinal polarization of the vector field.
 
\textit{Notations.} In this work, we use natural units with $\hbar=c=1$ and the reduced Planck mass $M_{\mathrm{P}} = 2.43\times 10^{18}\,\text{GeV}$. The inflationary background is described by the spatially flat Friedmann--Lema\^{i}tre--Robertson--Walker (FLRW) metric. Expressed in terms of physical time $t$ or conformal time $\eta$, the line element reads
\begin{equation}
\label{eq:FLRW-metric}
    ds^2 \!=\! g_{\mu\nu}\,\mathrm{d}x^\mu \mathrm{d}x^\nu \!=\! \mathrm{d}t^2\! -\! a^2(t)\,\mathrm{d}\bm{x}^2\! =\! a^2(\eta)(\mathrm{d}\eta^2\! -\! \mathrm{d}\bm{x}^2),
\end{equation}
where $a(t)$ is the scale factor. Derivatives with respect to cosmic time $t$ are denoted by overdots, while derivatives with respect to conformal time  $\eta$  are denoted by primes. The Hubble parameter is defined as $H \equiv \dot{a}/a = a'/a^2$.
%%%%%%%%%%%%%%%%%%%%%%%%%%%%%%%%%%%%%%%%%%%%%%%%%%%%%%%%%%%%%%%%%%%%%%%%%%

%%%%%%%%%%%%%%%%%%%%%%%%%%%%%%%%%%%%%%%%%%%%%%%%%%%%%%%%%%%%%%%%%%%%%%%%%%
\section{Model of the vector dark matter}
\label{sec:model}
The action that describes the dark vector field $A_{\mu}$ and the inflaton $\phi$, interacting through the kinetic and axial couplings, has the form
\begin{align}
\label{eq:action}
    &S[\phi,A_\mu]=\int d^{4}x\sqrt{-g} \Big[\frac{1}{2} \partial_{\mu}\phi\,\partial^{\mu}\phi\, - V(\phi) \\
    &-\frac{1}{4}I_{1}(\phi)F_{\mu\nu}F^{\mu\nu}-\frac{1}{4}I_{2}(\phi)F_{\mu\nu}\tilde{F}^{\mu\nu} +\frac{m^{2}(\phi)}{2}A_{\mu}A^{\mu} \Big]\,,\nonumber
\end{align}
where $g=\operatorname{det}(g_{\mu\nu})$ is the determinant of the spacetime metric, $V(\phi)$ is the inflaton potential, $I_{1}(\phi)$ is the kinetic (dilatonic) coupling function, $I_{2}(\phi)$ is the axial coupling function, $m(\phi)$ is the effective mass of the dark photon, $F_{\mu\nu}=\partial_{\mu}A_{\nu}-\partial_{\nu}A_{\mu}$ is the dark vector field tensor, $\tilde{F}^{\mu\nu}=
\varepsilon^{\mu\nu\sigma\rho}{F}_{\sigma\rho}/(2\sqrt{-g})$ is the corresponding dual tensor (here $\varepsilon^{\mu\nu\sigma\rho}$ is the totally antisymmetric Levi-Civita symbol with $\varepsilon^{0123} = +1$). The dual tensor satisfies the Bianchi identities
\begin{equation}
\label{eq:Bianchi-id}
    \frac{1}{\sqrt{-g}}\partial_{\mu}\left[\sqrt{-g}\,\tilde{F}^{\mu\nu} \right] = 0\,.
\end{equation} 

From action \eqref{eq:action}, we get the equations of motion for the inflaton field $\phi$
\begin{align}
\label{eq:KGF-cov}
    &\frac{1}{\sqrt{-g}}\partial_{\mu} \left[\sqrt{-g}\, \partial^{\mu}\phi \right] + \frac{dV}{d\phi}\\
    &=-\frac{1}{4}\frac{dI_{1}}{d\phi}F_{\mu\nu}F^{\mu\nu} - \frac{1}{4}\frac{dI_{2}}{d\phi}F_{\mu\nu}\tilde{F}^{\mu\nu} +m\frac{dm}{d\phi}A_{\mu}A^{\mu}=0\,.\nonumber
\end{align}
and for the gauge field
\begin{equation}
\label{eq:Proca-cov}
    \frac{1}{\sqrt{-g}}\partial_{\mu}\left[\sqrt{-g}\,I_{1}(\phi)F^{\mu\nu} \right]+ \frac{dI_{2}}{d\phi}\tilde{F}^{\mu\nu}\,\partial_{\mu}\phi+m^{2}A^{\nu}=0\,.
\end{equation}
Equation \eqref{eq:Proca-cov} together with the Bianchi identity \eqref{eq:Bianchi-id} forms the system of Proca equations for a massive vector field in the presence of kinetic and axial couplings in a curved spacetime. By varying the action \eqref{eq:action} with respect to the metric, we obtain the stress--energy tensor
\begin{align}
\label{eq:EMT-general}
    &T_{\mu\nu}\!=\!\frac{2}{\sqrt{-g}}\,\frac{\delta S}{\delta g^{\mu\nu}}\!=\!\partial_{\mu}\phi\,\partial_{\nu}\phi- I_{1}(\phi)F_{\mu\lambda}F_{\nu}^{\,\,\lambda}\,+m^{2}A_{\mu}A_{\nu}\nonumber\\
    &-g_{\mu\nu}\!\left[\frac{1}{2}\partial_{\alpha}\phi\,\partial^{\alpha}\phi  - \!V(\phi)\! -\! \frac{1}{4}I_{1}(\phi)F_{\alpha\beta}F^{\alpha\beta} +\frac{m^{2}}{2}A_{\alpha}A^{\alpha}\right].
\end{align}

Considering that $A^{\mu}=(A^{0}, a^{-2}\boldsymbol{A})$, we define the dark electric and magnetic fields%
\footnote{For convenience, we refer to the components of the field tensor of the dark massive vector field as the ``electric'' and ``magnetic'' fields by analogy with the ordinary Maxwell tensor in the standard electromagnetism.}
as
\begin{equation}
\label{E_and_B}
    \boldsymbol{E}=-\frac{1}{a}\left( \dot{\boldsymbol{A}}+\boldsymbol{\nabla}A^{0}\right), \quad \text{and} \quad \boldsymbol{B}=\frac{1}{a^{2}}\,{\rm rot}\boldsymbol{A}.
\end{equation}
The field tensor $F_{\mu\nu}$ and its dual $\tilde{F}_{\mu\nu}$ are expressed in terms of electric and magnetic fields as follows:
\begin{align}
\label{F}
    &F_{0i}=aE^{i},\qquad &F_{ij}=-a^{2}\epsilon_{ijk}B^{k},\nonumber\\
    &\tilde{F}^{0i}=-a^{-1}B^{i},\qquad&\tilde{F}^{ij}=a^{-2}\epsilon_{ijk}E^{k}\,.
\end{align}
     
Now, assuming the homogeneous and isotropic FLRW universe and spatially homogeneous inflaton field $\phi=\phi(t)$, we get the Friedmann equation governing the Universe expansion in the following form:
\begin{align}
\label{eq:Friedmann-explicit}
	&3H^{2}M_{\mathrm{P}}^{2}=\rho_{\mathrm{tot}}=\frac{1}{2}\dot{\phi}^{2}+V(\phi)\nonumber\\
    &\hspace{1cm} + \frac{I_{1}}{2}\langle \boldsymbol{E}^{2}+\boldsymbol{B}^{2}\rangle +\frac{m^{2}}{2}\langle A_{0}^{2}\rangle+\frac{m^{2}}{2a^{2}}\langle\boldsymbol{A}^{2}\rangle\, ,\\
\label{eq:Friedmann-2-explicit}
	&-(2\dot{H}+3H^{2})M_{\mathrm{P}}^{2}=p_{\mathrm{tot}}=\frac{1}{2}\dot{\phi}^{2}-V(\phi)\nonumber\\
    &\hspace{1cm} + \frac{I_{1}}{6}\langle \boldsymbol{E}^{2}+\boldsymbol{B}^{2}\rangle +\frac{m^{2}}{2}\langle A_{0}^{2}\rangle-\frac{m^{2}}{6a^{2}}\langle\boldsymbol{A}^{2}\rangle.
\end{align}
Here, the angular brackets denote the vacuum averaging of the operators of the electric and magnetic fields, scalar $A_{0}$ and vector $\boldsymbol{A}$ potentials.

The Klein--Gordon equation for the inflaton field reads
\begin{align}
\label{eq:KG}
    \ddot{\phi}+3H\dot{\phi}+\frac{dV}{d \phi}=&\frac{1}{2}\,\frac{dI_{1}}{d\phi}\langle \boldsymbol{E}^{2}-\boldsymbol{B}^{2} \rangle  + \frac{dI_{2}}{d\phi}\langle \boldsymbol{E}\cdot\boldsymbol{B} \rangle -\nonumber\\&-\frac{m}{a^{2}}\,\frac{dm}{d\phi}\langle\boldsymbol{A}^{2}\rangle+m\frac{dm}{d\phi}\langle A_{0}^{2}\rangle.
\end{align}
 
Finally, the Proca equations in vector form are
\begin{align}
\label{Proca_1}
    \dot{\boldsymbol{E}}+ 2 H\boldsymbol{E}- \frac{1}{a} \operatorname{rot} \boldsymbol{B}  +\frac{\dot{I}_1}{I_{1}}\boldsymbol{E} +\frac{\dot{I}_2}{I_{1}} \boldsymbol{B} - \frac{m^{2}}{a\, I_{1}}\boldsymbol{A}=0\,,\\
\label{Proca_2}
    \dot{\boldsymbol{B}}+2 H \boldsymbol{B}+\frac{1}{a}\operatorname{rot}\boldsymbol{E}=0\,, \\
\label{Proca_34}
    \operatorname{div} \boldsymbol{E}=-\frac{m^{2}\,a}{I_{1}}A_{0},\qquad\operatorname{div} \boldsymbol{B}=0\, .
\end{align}

For further convenience, let us decompose the vector potential, electric and magnetic fields into longitudinal and transverse components
\begin{equation}
\label{decompos}
    \boldsymbol{A}=\boldsymbol{A}_{\perp}+\boldsymbol{A}_{\parallel},\quad \boldsymbol{E}=\boldsymbol{E}_{\perp}+\boldsymbol{E}_{\parallel},\quad \boldsymbol{B}=\boldsymbol{B}_{\perp}.
\end{equation}
The longitudinal parts satisfy
\begin{equation}
    \operatorname{rot}\boldsymbol{A}_{\parallel} = 0\, , \qquad \operatorname{rot}\boldsymbol{E}_{\parallel} = 0\, ,
\end{equation}
while the transverse ones are determined such as
\begin{equation}
    \operatorname{div}\boldsymbol{A}_{\perp} = 0\, , \qquad \operatorname{div}\boldsymbol{E}_{\perp} = 0\, , \qquad\operatorname{div}\boldsymbol{B}_{\perp} = 0\, .
\end{equation}
In the next two sections, we study the evolution of the transverse and longitudinal quantities separately.

%%%%%%%%%%%%%%%%%%%%%%%%%%%%%%%%%%%%%%%%%%%%%%%%%%%%%%%%%%%%%%%%%%%%%%%%%%

%%%%%%%%%%%%%%%%%%%%%%%%%%%%%%%%%%%%%%%%%%%%%%%%%%%%%%%%%%%%%%%%%%%%%%%%%%
\section{Transverse polarizations of the vector field}
\label{sec:transverse}

\subsection{Mode evolution in Fourier space}

Let us first consider the transverse components of the vector field. In position space, they satisfy the transverse part of the Proca equations \eqref{Proca_1}--\eqref{Proca_2} while Eqs.~\eqref{Proca_34} are satisfied identically. Substituting $\boldsymbol{E}_\perp=-a^{-1} \dot{\boldsymbol{A}}_\perp$ and $\boldsymbol{B}_\perp=a^{-2}\operatorname{rot}\boldsymbol{A}_\perp$ in Eq.~\eqref{Proca_1} yields
\begin{equation}
\label{eq:EOM-vect-perp}
    \ddot{\boldsymbol{A}}_\perp+\Big(H+\frac{\dot{I}_1}{I_{1}}\Big)\dot{\boldsymbol{A}}_\perp-\frac{1}{a^2} \triangle\boldsymbol{A}_\perp - \frac{1}{a}\frac{\dot{I}_2}{I_{1}}{\rm rot}\boldsymbol{A}_\perp +\frac{m^{2}}{I_{1}}\boldsymbol{A}_\perp=0\,,
\end{equation}
where $\triangle=\partial_i \partial_i$ denotes the spatial Laplace operator.

The corresponding quantum field operator is decomposed over the set of Fourier modes in a standard way:
\begin{align}
\label{eq:quant-A-perp}
    \hat{A}_{\perp j}(t,\boldsymbol{x})=& \!\int\!\!\!\frac{d^{3}\boldsymbol{k}}{(2\pi)^{3/2}\sqrt{I_{1}}}\!\!\sum_{\,\lambda=\pm 1}\!\Big[\epsilon_{j}^{\lambda}(\boldsymbol{k})\hat{b}_{\boldsymbol{k},\lambda}\,\mathcal{A}_{\lambda}(t,k)e^{i\boldsymbol{k}\cdot\boldsymbol{x}}\nonumber\\
    &+\epsilon_{j}^{\lambda\ast}(\boldsymbol{k})\hat{b}_{\boldsymbol{k},\lambda}^{\dagger}\,\mathcal{A}_{\lambda}^{*}(t,k)e^{-i\boldsymbol{k}\cdot\boldsymbol{x}} \Big]\, ,
\end{align}
where $\mathcal{A}_{\lambda}(t,k)$ is the mode function; $\boldsymbol{\epsilon}^{\lambda}(\boldsymbol{k})$ are the polarization vectors of circular polarization $\lambda=\pm$ satisfying the following properties:
\begin{align}
\label{eq:polarization-vectors-prop}
    &\boldsymbol{\epsilon}^{\lambda}(\boldsymbol{k})\cdot\boldsymbol{k}=0,\qquad  \boldsymbol{\epsilon}^{\lambda\ast}(\boldsymbol{k})=\boldsymbol{\epsilon}^{\lambda}(-\boldsymbol{k})=\boldsymbol{\epsilon}^{-\lambda}(\boldsymbol{k}),\\ &\boldsymbol{\epsilon}^{\lambda\ast}(\boldsymbol{k})\cdot\boldsymbol{\epsilon}^{\lambda'}(\boldsymbol{k})=\delta^{\lambda\lambda'},\qquad
    i\boldsymbol{k}\times\boldsymbol{\epsilon}^{\lambda}(\boldsymbol{k}) = \lambda k \boldsymbol{\epsilon}^{\lambda}(\boldsymbol{k}),\\
    &\sum_{\lambda=\pm 1}\epsilon^{\lambda}_{i}(\boldsymbol{k})\epsilon^{\lambda\ast}_{j}(\boldsymbol{k})=\delta_{ij}-\dfrac{k_{i}k_{j}}{k^{2}}\, ;
\end{align}
and $\hat{b}_{\boldsymbol{k},\lambda}$ ($\hat{b}^\dagger_{\boldsymbol{k},\lambda}$) denote the annihilation (creation) operators satisfying the canonical commutation relations
\begin{equation}
\label{eq:commut-perp}
    \big[\hat{b}^{\vphantom{\dagger}}_{\boldsymbol{k},\lambda},\ \hat{b}_{\boldsymbol{k}',\lambda'}^{\dagger}  \big]=\delta^{(3)}\!\big(\boldsymbol{k}-\boldsymbol{k}'\big)\,\delta_{\lambda\lambda'}\,.
\end{equation}
Then, from Eq.~\eqref{eq:EOM-vect-perp}, we get the following equation for the mode function:   
\begin{align}
\label{eq:EOM-A-lambda}
    \ddot{\mathcal{A}}_{\lambda}+H\dot{\mathcal{A}}_{\lambda}+&\Big[\frac{k^{2}}{a^{2}}+\frac{m^2}{I_1}-\lambda\frac{k}{a}\,\frac{\dot{I_{2}}}{I_{1}}\nonumber\\
    &-\frac{\ddot{I}_1+H\dot{I}_1}{2I_1}+\Big(\frac{\dot{I_{1}}}{2I_{1}}\Big)^2\Big] \mathcal{A}_{\lambda} =0\, .
\end{align}

Introducing the function
\begin{equation}
\label{D_lambda}
    \mathcal{D}_{\lambda}(t,k)=\frac{\sqrt{I_{1}}\,a}{k}\frac{\partial}{\partial t} \left( \frac{\mathcal{A}_{\lambda}}{\sqrt{I_{1}}} \right) =\frac{\sqrt{I_{1}}}{k}\frac{\partial}{\partial \eta} \left( \frac{\mathcal{A}_{\lambda}}{\sqrt{I_{1}}}\right)\,,
\end{equation}
one can  represent Eq.~\eqref{eq:EOM-A-lambda} in an equivalent form of two first-order ordinary differential equations
\begin{align}
    \dot{\mathcal{A}}_\lambda &= \frac{\dot{I}_1}{2I_1}\mathcal{A}_\lambda+\frac{k}{a}\mathcal{D}_\lambda \, ,\\
    \dot{\mathcal{D}}_\lambda &= -\frac{\dot{I}_1}{2I_1}\mathcal{D}_\lambda-\Big(\frac{k}{a}-\lambda\frac{\dot{I}_2}{I_1}+\frac{m^2 a}{kI_1}\Big)\mathcal{A}_\lambda \, ,
\end{align}
which are particularly useful for numerical  computations.

For further convenience, we introduce the following notations:
\begin{equation}
\label{eq:gamma-def}
    \gamma(t)=\frac{\dot{I}_{1}}{2HI_{1}}\, ,
\end{equation}
\begin{equation}
\label{eq:xi-def}
    \xi(t)=\frac{\dot{I}_{2}}{2HI_{1}}\, ,
\end{equation}
\begin{equation}
\label{eq:mu-def}
    \mu(t)=\frac{m}{H\sqrt{I_{1}}}\, ,
\end{equation}
\begin{align}
\label{eq:S-def}
    S(t)&=\frac{\ddot{I}_1+H\dot{I}_1}{2I_1 H^{2}}-\Big(\frac{\dot{I}_1}{2HI_{1}}\Big)^2-\frac{m^2}{I_1 H^2}\nonumber\\
    &=\gamma^2 + \gamma(1-\epsilon_H-\epsilon_\gamma)-\mu^2\, ,
\end{align}
where $\epsilon_x \equiv -\dot{x}/(xH)$ represents the slow-roll parameter for the quantity $x$.

Switching to the new variable $z=k\eta$, where $\eta$ is the conformal time, Eq.~\eqref{eq:EOM-A-lambda} can be rewritten as
\begin{equation}
\label{eq:EOM-A-lambda-z1}
    \frac{\partial^{2}\!\mathcal{A}_{\lambda}}{\partial z^{2}}+\bigg[1-\frac{2\lambda\xi(t)}{(k/aH)}-\frac{S(t)}{(k/aH)^{2}} \bigg]\mathcal{A}_{\lambda}=0\,.
\end{equation}
For the Fourier mode deep inside the horizon, $k/aH\gg 1$ (corresponding to $-z\gg 1$ during inflation), the first term in brackets in Eq.~\eqref{eq:EOM-A-lambda-z1} dominates and the equation reduces to that of a harmonic oscillator with unit frequency. To select only the positive-frequency solution to that equation, we impose the Bunch--Davies vacuum boundary condition
\begin{equation}
\label{F_BD}
    \mathcal{A}_{\lambda}\simeq \frac{1}{\sqrt{2k}}\,e^{-iz}, \qquad -z\gg 1.
\end{equation}
This fully specifies the solution of the mode equation for all subsequent times. 

As time evolves, the other two terms in brackets become more and more important. Finally, at some point, they become comparable to unity. In the physically interesting situation, the whole expression in brackets can cross zero, and the mode becomes tachyonically unstable. We \textit{define} the momentum of the mode that crosses the horizon in the following way:
\begin{equation}
\label{eq:kh-def}
    k_h(t) = \underset{t'\leq t}{\operatorname{max}}\big[a(t')H(t') r(t')\big]\, ,
\end{equation}
where 
\begin{equation}
\label{eq:r-function}
    r(t)=|\xi(t)|+\sqrt{\xi^2(t)+|S(t)|}\, ,
\end{equation}
and we take the maximal value of the expression in Eq.~\eqref{eq:kh-def} to account for its possible nonmonotonic behavior in realistic models. For any given momentum $k$, one can find a moment of time $t_h(k)$ when it crosses the horizon, i.e.,
\begin{equation}
    t_h(k)=\underset{i}{\operatorname{min}}\{t_i:\, a(t_i)H(t_i)r(t_i)=k\}\, .
\end{equation}
The moment of time $t_h(k)$ is very special for the mode with momentum $k$ because around it, the mode changes its behavior qualitatively from fast oscillations corresponding to vacuum fluctuations to a smooth evolution with possible tachyonic enhancement. 

For the GEF in the next subsection, we will need the mode function at the moment of horizon crossing $t_h$. To find it, we assume that (i)~the quantities $\xi$ and $S$ in Eq.~\eqref{eq:EOM-A-lambda-z1} are smooth and slowly varying during inflation and (ii)~the Universe expansion is quasiexponential. The first assumption allows us to set $\xi$ and $S$ constant while looking for the solution close to the horizon crossing and the second assumption enables us to use the de Sitter solution $\eta \approx -1/(aH)$. Then, Eq.~\eqref{eq:EOM-A-lambda-z1} takes the form
\begin{equation}
\label{eq:EOM-A-lambda-z2}
    \frac{\partial^{2}\!\mathcal{A}_{\lambda}}{\partial z^{2}}+\bigg[1+\frac{2\lambda\xi(t_{h})}{z}-\frac{S(t_{h})}{z^{2}} \bigg]\mathcal{A}_{\lambda}=0\, ,
\end{equation}
which is the Whittaker equation. Its solution satisfying the Bunch--Davies boundary condition is expressed in terms of the Whittaker $W$ function:
\begin{equation}
\label{A_Wittaker}
    \mathcal{A}_{\lambda}(z,k)=\frac{1}{\sqrt{2k}}\,e^{\lambda\pi\xi(t_{h})/2}\,W_{\kappa,\, \nu}(2iz)\, ,
\end{equation}
where
\begin{equation}
    \kappa=-i\lambda\xi(t_{h}),  \qquad  \nu=\sqrt{\frac{1}{4}+S(t_{h})}.
\end{equation}
The corresponding expression for the function $\mathcal{D}_{\lambda}$ reads
\begin{align}
\label{D_Wittaker}
    \mathcal{D}_{\lambda}(z,k)=\frac{1}{\sqrt{2k}}\frac{e^{\lambda\pi\xi/2}}{z}\Big[&(iz+i\lambda\xi +\gamma) W_{-i\lambda\xi,\,\nu}(2iz) \nonumber\\
    &-\,W_{1-i\lambda\xi,\,\nu}(2iz)\Big]\, ,
\end{align}
where the parameters $\xi$, $\gamma$, $\nu$ are computed at the moment of horizon crossing, $t_h(k)$.

\subsection{Gradient-expansion formalism}

If the produced gauge field is strong enough so that its backreaction becomes nonnegligible, the dynamics of the system becomes highly nonlinear and all Fourier modes are coupled. In this case, it is more convenient to keep working in the coordinate space and introduce the following set of bilinear quantities:
\begin{align}
	\label{E_n}
	\mathcal{E}^{(n)}&=\phantom{-}\frac{I_{1}}{a^{n}}\langle \boldsymbol{E}_{\perp}\cdot \operatorname{rot}^{n} \boldsymbol{E}_{\perp}  \rangle,\\ 
	\label{G_n}
	\mathcal{G}^{(n)}&=-\frac{I_{1}}{2a^{n}}\langle \boldsymbol{E}_{\perp}\cdot \operatorname{rot}^{n} \boldsymbol{B}_{\perp}  + \operatorname{rot}^{n} \boldsymbol{B}_{\perp}\cdot \boldsymbol{E}_{\perp}  \rangle,\\
	\label{B_n}
	\mathcal{B}^{(n)}&=\phantom{-}\frac{I_{1}}{a^{n}}\langle \boldsymbol{B}_{\perp}\cdot\operatorname{rot}^{n} \boldsymbol{B}_{\perp}  \rangle.
\end{align}
Here the angular brackets denote the vacuum expectation value and $n$ is an integer number that can take values from $0$ (for $\mathcal{E}^{(n)}$), $-1$ (for $\mathcal{G}^{(n)}$), or $-2$ (for $\mathcal{B}^{(n)}$) to arbitrarily large positive numbers. Negative power of the curl operator has to be understood as an inverse operator; e.g., from Eq.~\eqref{E_and_B}, we get
\begin{equation}
\label{eq:rot-inverse-B}
    \operatorname{rot}^{-1}\boldsymbol{B}_\perp = \frac{1}{a^2}\boldsymbol{A}_\perp\,.
\end{equation}
This relation gives a unique result in the transverse sector. Using Eq.~\eqref{eq:rot-inverse-B}, we can easily get the explicit expressions for the bilinear quantities with negative indices,
\begin{align}
\mathcal{G}^{(-1)} &= -\frac{I_{1}}{2a}\langle \boldsymbol{E}_{\perp}\cdot  \boldsymbol{A}_{\perp}  + \boldsymbol{A}_{\perp}\cdot \boldsymbol{E}_{\perp}  \rangle\, ,\\
\mathcal{B}^{(-1)} &= \phantom{-}\frac{I_{1}}{a}\langle \boldsymbol{B}_{\perp}\cdot \boldsymbol{A}_{\perp}  \rangle \equiv \mathcal{H}_{B}\, ,\\
\mathcal{B}^{(-2)} &= \phantom{-}\frac{I_{1}}{a^2}\langle \boldsymbol{A}_{\perp}^2\rangle\, .
\end{align}
Here $\mathcal{H}_{B}$ is the magnetic helicity.

Now, using the Proca equations \eqref{Proca_1}--\eqref{Proca_34}, it is straightforward to derive equations of motion for the bilinear quantities \eqref{E_n}--\eqref{B_n},
\begin{align}
\label{dot_E_n}
    &\dot{\mathcal{E}}^{(n)} + (n+4+2\gamma)H \mathcal{E}^{(n)} - 4\xi H \mathcal{G}^{(n)} +2\mathcal{G}^{(n+1)}\nonumber\\
    &\hspace{2cm}+2\frac{m^{2}}{I_{1}}\,\mathcal{G}^{(n-1)}=[\dot{\mathcal{E}}^{(n)}]_{\rm b}\,,\\
\label{dot_G_n}
    &\dot{\mathcal{G}}^{(n)} +(n+4)H \,\mathcal{G}^{(n)}- 2\xi H  \mathcal{B}^{(n)}-\mathcal{E}^{(n+1)}+\mathcal{B}^{(n+1)} \nonumber\\
    &\hspace{2cm}+\frac{m^{2}}{I_{1}}\,\mathcal{B}^{(n-1)}=[\dot{\mathcal{G}}^{(n)}]_{\rm b}\,,\\
\label{dot_B_n}
    &\dot{\mathcal{B}}^{(n)} + (n+4-2\gamma)H	\mathcal{B}^{(n)}-2\mathcal{G}^{(n+1)} =[\dot{\mathcal{B}}^{(n)}]_{\rm b}\,.
\end{align}
These equations are supplemented by extra terms on the right-hand side, the so-called boundary terms, which originate from the contributions of vacuum gauge-field modes crossing the horizon. They act like a quantum source pumping the energy from vacuum modes and, by this, at least partially counteract the redshift due to the Universe's expansion.

In order to better understand the reason for the occurrence of boundary terms and derive explicit expressions for them, we need to find the spectral representations for the quantities $\mathcal{E}^{(n)}, \mathcal{G}^{(n)}$, and $\mathcal{B}^{(n)}$. Using Eq.~\eqref{eq:quant-A-perp}, as well as \eqref{E_n}--\eqref{B_n}, we obtain the following expressions:
\begin{align}
\label{E_1}
    \mathcal{E}^{(n)}&=\sum_{\lambda=\pm 1}\int_{0}^{k_{h}}\frac{d k}{k} \lambda^{n}\frac{k^{n+3}}{2\pi^{2}a^{n+2}}\,I_{1}\,\bigg| \frac{\partial}{\partial t}\left(\frac{\mathcal{A}_{\lambda}}{\sqrt{I_{1}}} \right) \bigg| ^{2} \nonumber\\
    &=\sum_{\lambda=\pm 1}\int_{0}^{k_{h}}\frac{d k}{k} \lambda^{n}\frac{k^{n+5}}{2\pi^{2}a^{n+4}}|\,\mathcal{D}_{\lambda}|^{2}\, ,\\
\label{G_1}
    \mathcal{G}^{(n)}&=\sum_{\lambda=\pm 1}\int_{0}^{k_{h}}\frac{d k}{k} \lambda^{n+1}\frac{k^{n+4}}{4\pi^{2}a^{n+3}}\,I_{1}\,\frac{\partial}{\partial t}\bigg|\frac{\mathcal{A}_{\lambda}}{\sqrt{I_{1}}}\bigg|^{2}\nonumber\\
    &=\sum_{\lambda=\pm 1}\int_{0}^{k_{h}}\frac{d k}{k} \lambda^{n+1}\frac{k^{n+5}}{2\pi^{2}a^{n+4}}\operatorname{Re}\left[\mathcal{A}_{\lambda}^{*}\mathcal{D}_{\lambda} \right]\, , \\
\label{B_1}
    \mathcal{B}^{(n)}&=\sum_{\lambda=\pm 1}\int_{0}^{k_{h}}\frac{d k}{k} \lambda^{n}\frac{k^{n+5}}{2\pi^{2}a^{n+4}}\,|\mathcal{A}_{\lambda}|^{2}\,.
\end{align}
Here, the upper integration boundary is replaced by a finite momentum. By this, we regularize the UV divergent integrals for the bilinear quantities. Although this is far from a rigorous renormalization (as it is done in quantum field theory in Minkowski space), this trick allows us to separate the relevant Fourier modes and disregard the contributions of vacuumlike modes.

Therefore, any gauge-field quantity $X$ can be represented in the form
\begin{equation}
\label{X_integr}
    X=\int_{0}^{k_{h}}\frac{dk}{k}\frac{dX}{d\ln k},
\end{equation}
where $k_{h}=k_{h}(t)$ is the wave number of the mode crossing the horizon at the given time $t$, see Eq.~\eqref{eq:kh-def} for its explicit form. The value of $X$ depends on time for two reasons. First, its spectral density $dX/d\ln k$ changes over time. And second, the integral over momentum has a variable upper integration boundary. Terms on the left-hand side in Eqs.~\eqref{dot_E_n}--\eqref{dot_B_n} take into account the first time dependence. In order to take care of the second one, we need to introduce the boundary terms,
\begin{equation}
\label{X_p_d}
    (\dot{X})_{\rm b}= \frac{d \ln k_{h}}{d t}\cdot\left. \frac{dX}{d\ln\, k}  \right|_{k=k_h(t)}.
\end{equation}
To calculate this boundary term, we need to know the spectral density of the quantity $X$ at the moment the mode crosses the horizon. For this, we use the explicit expressions for the mode functions $\mathcal{A}_\lambda$ and $\mathcal{D}_\lambda$ given in Eqs.~\eqref{A_Wittaker} and \eqref{D_Wittaker}, which are valid close to the moment of horizon crossing. 

The momentum of the mode that is crossing the horizon at the moment of time $t$ is represented as an upper envelope of the expression $aHr$, where $r$ is given in Eq.~\eqref{eq:r-function}. From the definition of the upper monotonic envelope, it follows that when $k_h(t)$ increases in time, it just coincides with the value of $a(t)H(t)r(t)$. When $k_h(t)$ is not equal to $a(t)H(t)r(t)$, it is just constant, i.e., $d\ln k_h/dt=0$. In the latter case, the boundary term just vanishes. Therefore, when substituting $k_h$ to the expressions for the mode functions $\mathcal{A}_\lambda$ and $\mathcal{D}_\lambda$ we may simply use $k_h(t)=a(t)H(t)r(t)$.

Now, using Eq.~\ref{X_p_d}, we obtain the expressions for the boundary terms,
\begin{align}
\label{E_p_d}
    &[\dot{\mathcal{E}}^{(n)}]_{\rm b}=\frac{d \ln k_{h}}{d t}\,\frac{H^{n+4}r^{n+2}}{4\pi^{2}}\sum_{\lambda=\pm 1}\!\lambda^{\!n}\,e^{\lambda\pi\xi}\nonumber\\
    &\times \bigg|\Big(ir-i\lambda\xi-\gamma  \Big)W_{-i\lambda\xi, \,\nu}(-2ir) +W_{1-i\lambda\xi, \,\nu}(-2ir)  \bigg| ^{2},\\
\label{G_p_d}
    &[\dot{\mathcal{G}}^{(n)}]_{\rm b}=\frac{d \ln k_{h}}{d t}\,\frac{H^{n+4}r^{n+3}}{4\pi^{2}}\sum_{\lambda=\pm 1}\!\lambda^{\!n+1}\,e^{\lambda\pi\xi}\nonumber\\
    &\times \bigg\{\! \mathrm{Re}\left[W_{i\lambda\xi, \,\nu}(2ir)W_{1-i\lambda\xi, \,\nu}(-2ir) \right]\! -\!\gamma \left|W_{i\lambda\xi, \,\nu}(2ir) \right|^{2}\!\bigg\},\\
\label{B_p_d}
    &[\dot{\mathcal{B}}^{(n)}]_{\rm b}=\frac{d \ln k_{h}}{d t}\,\frac{H^{n+4}r^{n+4}}{4\pi^{2}}\sum_{\lambda=\pm 1}\!\lambda^{\!n}\,e^{\lambda\pi\xi}\,\, \big|W_{i\lambda\xi, \,\nu}(2ir) \big|^{2}.
\end{align}

Note that the system of equations \eqref{dot_E_n}--\eqref{dot_B_n} is, in principle, infinite. Indeed, the equation of motion for the quantity of order $n$  contains the quantity of the order $(n+1)$. In practice, this system has to be truncated at some finite order $n_{\mathrm{max}}$. For this, one needs to express the quantities of the order $(n_{\mathrm{max}}+1)$ in terms of lower-order quantities.
Fortunately, the truncation condition can be easily guessed from the spectral decompositions \eqref{E_1}--\eqref{B_1}. Indeed, for large $n$, the integral over the momentum is dominated by the upper integration boundary. Therefore, one can approximately write
\begin{equation}
\label{eq:truncation-transverse}
    \mathcal{E}^{(n_{\mathrm{max}}+1)}\approx \Big(\frac{k_h}{a}\Big)^2 \mathcal{E}^{(n_{\mathrm{max}}-1)}
\end{equation}
and similar relations for $\mathcal{G}^{(n_{\mathrm{max}}+1)}$ and $\mathcal{B}^{(n_{\mathrm{max}}+1)}$. Note that more complicated truncation conditions, like the one proposed in Ref.~\cite{Domcke:2023tnn}, can be used as well. In practice, the required truncation order is determined numerically: we increase the order $n_{\mathrm{max}}$ until the solution converges and no longer changes with further increases of $n_{\mathrm{max}}$.

Despite the fact that Eqs.~\eqref{dot_E_n} and \eqref{dot_G_n} for $n$th-order quantities also contain the terms of the order $(n-1)$, the system is not infinite in the direction of negative orders. Indeed, Eq.~\eqref{dot_B_n} does not contain such lower-order terms; this allows one to close the system. Only three negative-order quantities are relevant: $\mathcal{G}^{(-1)}$, $\mathcal{B}^{(-1)}$, and $\mathcal{B}^{(-2)}$.

In the end of this section, we would like to note that the GEF for the transverse vector-field modes presented here is very similar to the one developed for the massless gauge field in Refs.~\cite{Sobol:2020lec,Gorbar:2021rlt,Durrer:2023rhc,Lysenko:2025pse} with the only differences coming from the presence of the mass term. In the limit $\mu = m/(H\sqrt{I_1})\ll 1$ (which is a typical situation for realistic models of the dark photon), this formalism simply reduces to the one reported in Refs.~\cite{Durrer:2023rhc,Lysenko:2025pse}. This will not be the case for longitudinal modes which we consider in the next section.

%%%%%%%%%%%%%%%%%%%%%%%%%%%%%%%%%%%%%%%%%%%%%%%%%%%%%%%%%%%%%%%%%%%%%%%%%%

%%%%%%%%%%%%%%%%%%%%%%%%%%%%%%%%%%%%%%%%%%%%%%%%%%%%%%%%%%%%%%%%%%%%%%%%%%
\section{Longitudinal polarizations of the vector field}
\label{sec:longitudinal}

\subsection{Mode evolution in Fourier space}

Now, we switch to the longitudinal modes of the gauge field. Their evolution is governed by the longitudinal part of Eq.~\eqref{Proca_1} and the first equation in \eqref{Proca_34}. The Faraday law \eqref{Proca_2} and the second equation in \eqref{Proca_34} are satisfied identically. Making use of $\boldsymbol{E}_{\parallel}=-(\dot{\boldsymbol{A}}_{\parallel}+\boldsymbol{\nabla}A_0)/a$ and $\boldsymbol{B}_\parallel =0$, we rewrite the two relevant equations in the following form:
\begin{align}
\label{eq:eom-long-1}
    &\ddot{\boldsymbol{A}}_\parallel+\Big(H+\frac{\dot{I}_1}{I_{1}}\Big)\dot{\boldsymbol{A}}_\parallel+\frac{m^{2}}{I_{1}}\boldsymbol{A}_\parallel=-\boldsymbol{\nabla}\Big[\dot{A}_0 + \Big(H+\frac{\dot{I}_1}{I_{1}}\Big)A_0\Big],\\
\label{eq:eom-long-2}
    &\hspace{2cm}\operatorname{div}\dot{\boldsymbol{A}}_\parallel = \Big(\frac{m^2 a^2}{I_1}-\triangle\Big)A_0\, .
\end{align}
These equations describe the dynamics of one physical degree of freedom\,---\,the longitudinal component of the gauge field. The component $A_0$ is not independent and can be determined from the known $\boldsymbol{A}_\parallel$ via Eq.~\eqref{eq:eom-long-2}. Considering the quantized fields, we decompose the field operators in coordinate space over the set of longitudinal Fourier modes as follows:
\begin{align}
\label{eq:quant-A-par}
	\hat{\boldsymbol{A}}_\parallel(t,\boldsymbol{x})=\!\!\int\!\!\! \frac{d^{3}\boldsymbol{k}}{(2\pi)^{3/2}\sqrt{I_{1}}\,m}\Big[\frac{\boldsymbol{k}}{k}\hat{b}^{\vphantom{\dagger}}_{\boldsymbol{k},L}\,\mathcal{A}^{\vphantom{\ast}}_{L}(t,k)e^{i\boldsymbol{k}\cdot\boldsymbol{x}}+\mathrm{h.c.}\Big],\\
\label{eq:quant-A-par0}
	\hat{A}_{0}(t,\boldsymbol{x})=\!\!\int\!\!\!\frac{d^{3}\boldsymbol{k}}{(2\pi)^{3/2}\sqrt{I_{1}}\,m}\Big[\frac{i}{a}\hat{b}^{\vphantom{\dagger}}_{\boldsymbol{k},L}\,\mathcal{D}^{\vphantom{\ast}}_{L}(t,k)e^{i\boldsymbol{k}\cdot\boldsymbol{x}}+\mathrm{h.c.}\Big],
\end{align}
where $\mathcal{A}_{L}(t,k)$ and $\mathcal{D}_{L}(t,k)$ are the mode functions, $\hat{b}^{\vphantom{\dagger}}_{\boldsymbol{k},L}$ ($\hat{b}^{\dagger}_{\boldsymbol{k},L}$) are the annihilation (creation) operators of the longitudinal mode with momentum $\boldsymbol{k}$, and ``h.c.'' denotes the Hermitian conjugate terms. The operators $\hat{b}^{\vphantom{\dagger}}_{\boldsymbol{k},L}$  and $\hat{b}^{\dagger}_{\boldsymbol{k},L}$ satisfy the canonical commutation relations
\begin{equation}
\label{eq:commut-par}
    \big[\hat{b}^{\vphantom{\dagger}}_{\boldsymbol{k},L},\ \hat{b}^{\dagger}_{\boldsymbol{k}',L}\big]=\delta^{(3)}\!\big(\boldsymbol{k}-\boldsymbol{k}'\big)\, .
\end{equation}
Note that they also commute with all annihilation and creation operators of the transverse modes.

Equation~\eqref{eq:eom-long-2} gives the relation between the mode functions $\mathcal{A}_{L}(t,k)$ and $\mathcal{D}_{L}(t,k)$,
\begin{equation}
\label{eq:eom-AL-system}
    \dot{\mathcal{A}}_{L}- \left(\frac{\dot{m}}{m}+\frac{\dot{I_{1}}}{2I_{1}}\right) \mathcal{A}_{L}-\Big(\frac{k}{a}+\frac{m^2}{I_1}\frac{a}{k}\Big)\mathcal{D}_{L}=0\, .
\end{equation}
Then, a cumbersome but rather straightforward computation leads from Eq.~\eqref{eq:eom-long-1} to the second equation of motion
\begin{equation}
\label{eq:eom-DL-system}
    \dot{\mathcal{D}}_{L}+\Bigg[2H+\frac{\dot{m}}{m}-\frac{\dot{I_{1}}}{2I_{1}}\Bigg]\mathcal{D}_{L}+\frac{k}{a}\mathcal{A}_{L}=0\, ,
\end{equation}
which constitutes, together with the previous one, the simplest system of equations suitable for the numerical implementation. In Appendix~\ref{app:longitudinal}, we derive the boundary conditions which supplement the system above; they are based on the assumption that the longitudinal polarization of the gauge field is in the Bunch--Davies vacuum state far inside the Hubble horizon. If the parameter $\mu=m/(\sqrt{I_1}H)\ll 1$ (which is the case for the absolute majority of the dark photon models), the boundary conditions take the form,
\begin{align}
\label{eq:BC-AL}
    \mathcal{A}_L(t,k)&\simeq \frac{1}{a}\sqrt{\frac{kI_1(t)}{2}}e^{-ik\eta(t)},\\
\label{eq:BC-DL}
    \mathcal{D}_L(t,k)&\simeq \sqrt{\frac{I_1(t)}{2k}}\Big(-i\frac{k}{a}-H-\frac{\dot{m}}{m}\Big)e^{-ik\eta(t)},
\end{align}
both valid as $-k\eta(t)\gg 1$. 

One can also derive a single second-order differential equation for the mode function $\mathcal{A}_{L}$, in full analogy with the transverse-polarizations case; cf. Eq.~\eqref{eq:EOM-A-lambda}. However, in this case, it takes a rather cumbersome form, which we give in Appendix~\ref{app:longitudinal} [see Eq.~\eqref{eq:eom-AL-1} there] not to distract the reader from the main flow of the paper. One can use this equation for numerical calculations as well. 

The only remaining thing that we would like to discuss here is the approximate behavior of the mode functions close to the horizon crossing. For this, we introduce the canonically normalized longitudinal mode function
\begin{equation}
\label{eq:canonical-mode-fn}
    \psi_L(t,k)=\Big(m^2+\frac{k^2 I_1}{a^2}\Big)^{-1/2}\mathcal{A}_L(t,k)\, ,
\end{equation}
whose equation of motion has the form
\begin{equation}
\label{eq:eom-psi-L}
    \ddot{\psi}_L + H\dot{\psi}_L + Q \psi_L=0\,
\end{equation}
where the function $Q=Q(t,k)$ is given by Eq.~\eqref{eq:Q-func} in Appendix~\ref{app:longitudinal}. Now, we will consider the mode close to the horizon crossing, i.e., $k/a\sim H$. In this case, using the assumption that $\mu\ll 1$, we can get a simplified form of the function $Q$ as follows:
\begin{equation}
\label{eq:Q-approx}
    Q\big\vert_{\frac{k}{aH}\gtrsim 1,\,\mu\ll 1} \approx \frac{k^2}{a^2}-\frac{\ddot{a}}{a}-\frac{\ddot{m}}{m}-H^2-3H\frac{\dot{m}}{m}\, .
\end{equation}
For convenience, we introduce a dimensionless parameter
\begin{equation}
\label{eq:M-def}
    M=\frac{1}{H}\frac{\dot{m}}{m}\, ,
\end{equation}
and its corresponding slow-roll parameter $\epsilon_M=-\dot{M}/(MH)$. Finally, we switch to the variable $z=k\eta(t)$ and obtain the canonical mode equation in the form:
\begin{equation}
\label{eq:mode-psi-L-z}
    \frac{\partial^2\psi_L}{\partial z^2}+\bigg[1-\frac{R^2}{(k/aH)^2}\bigg]\psi_L=0\,
\end{equation}
where
\begin{equation}
\label{eq:R-def}
    R^2(t)=2+3M+M^2-\epsilon_H(1+M)-M\epsilon_M\, .
\end{equation}
From Eq.~\eqref{eq:mode-psi-L-z} we can find a reasonable definition of the momentum of the mode that crosses the horizon. In fact, when the second term in brackets becomes comparable to unity, the behavior of the mode starts deviating from the vacuum solution. Therefore, we \textit{define}
\begin{equation}
\label{eq:khL-def}
    k_{h,L}(t) = \underset{t'\leq t}{\operatorname{max}}\big[a(t')H(t') |R(t')|\big]\, ,
\end{equation}
where again the upper monotonic envelope is used to ensure that once the mode crosses the horizon it never crosses it back during inflation. Further, for any given mode $k$, we can find a moment of time $t_{h,L}$ when it crosses the horizon,
\begin{equation}
\label{eq:thL-def}
    t_{h,L}=\underset{i}{\operatorname{min}}\{t_i\, :\, a(t_i)H(t_i)|R(t_i)|=k\}\, .
\end{equation}
Then, to get an approximate solution of Eq.~\eqref{eq:mode-psi-L-z} valid before and around the horizon crossing, we (i) assume the pure de Sitter expansion, $H=\text{const}$, $aH=-1/\eta$; and (ii) treat $R$ as constant by taking its value at the moment of horizon crossing (as deep inside the horizon, the mode does not feel $R$ at all). Equation~\eqref{eq:mode-psi-L-z} takes the form
\begin{equation}
\label{eq:eom-psiL-2}
    \frac{\partial^2 \psi_L}{\partial z^2} + \bigg[1-\frac{R^2(t_{h,L})}{z^2}\bigg]\psi_L=0\, .
\end{equation}
Its solution, satisfying the Bunch--Davies boundary condition~\cite{Bunch:1978yq}
\begin{equation}
\label{eq:Bunch-Davies-BC-L}
    \psi_L\simeq \frac{1}{\sqrt{2k}}e^{-iz}\,, \qquad -z\gg 1\,,
\end{equation}
is expressed in terms of the Whittaker function or the Hankel function of the first kind as follows:
\begin{align}
\label{eq:psi-L-solution}
    \psi_L(z,k)&=\frac{1}{\sqrt{2k}}W_{0,\alpha}(2iz)\nonumber\\
    &=\frac{e^{i\pi(2\alpha+1)/4}}{\sqrt{2k}}\sqrt{\frac{\pi}{2}(-z)}H^{(1)}_{\alpha}(-z)\, ,
\end{align}
where
\begin{equation}
\label{eq:alpha-def}
    \alpha=\sqrt{\frac{1}{4}+R^2(t_{h,L})}\, .
\end{equation}

Finally, the expressions for the mode functions $\mathcal{A}_L$ and $\mathcal{D}_L$ read
\begin{align}
    &\mathcal{A}_L(z,k)=\hphantom{-}e^{i\pi(2\alpha+1)/4}\frac{H}{2}\sqrt{\frac{\pi I_1}{k}} (-z)^{3/2} H^{(1)}_{\alpha}(-z)\, ,\\
    &\mathcal{D}_L(z,k)=-e^{i\pi(2\alpha+1)/4}\frac{H}{2}\sqrt{\frac{\pi I_1}{k}} (-z)^{3/2}\nonumber\\
    &\hspace{0.7cm}\times\bigg[\frac{3/2+\alpha+M}{(-z)}H^{(1)}_{\alpha}(-z)-H^{(1)}_{\alpha+1}(-z)\bigg]\, ,
\end{align}
where the parameters $M$, $\gamma$, and $\alpha$ are computed at the moment of horizon crossing, $t_{h,L}(k)$.

\subsection{Gradient-expansion formalism}

Let us introduce an auxiliary quantity 
\begin{equation}
    \Phi=\frac{1}{a}(-\triangle)^{-1/2}\operatorname{div}\boldsymbol{A}_{\parallel}.
\end{equation}
It has a similar spectral decomposition as $A_0$, Eq.~\eqref{eq:quant-A-par0}, with the replacement of the mode function $\mathcal{D}_L$ by $\mathcal{A}_L$. In this expression and in what follows, fractional and negative powers of the Laplacian operator have to be treated formally by keeping in mind the Fourier representation of the corresponding quantities.
From Eqs.~\eqref{eq:eom-long-1}--\eqref{eq:eom-long-2}, we get the following system of equations for two scalar functions, $A_0$ and $\Phi$:
\begin{align}
\label{eq:eom-A0}
	\dot{A}_0&=-H(3+2M)A_{0}-\frac{(-\triangle)^{1/2}}{a}\Phi,\\
\label{eq:eom-Phi}
	\dot{\Phi}&=-H\Phi+\frac{m^{2}}{I_{1}}\frac{a}{(-\triangle)^{1/2}}A_{0}+ \frac{(-\triangle)^{1/2}}{a} A_{0}.
\end{align}

With the aim of constructing the analog of the GEF from the previous section, we introduce the following set of bilinear quantities:
\begin{align}
\label{eq:Q_2p}
	\mathcal{Q}^{(2p)}&=\frac{m^{2}}{a^{2p}}\,\langle A_{0}\, (-\triangle)^{p} A_{0}  \rangle\, ,\\
\label{eq:F_2p}
	\mathcal{F}^{(2p)}&=\frac{m^{2}}{a^{2p}}\,\langle \Phi\, (-\triangle)^{p} \Phi   \rangle\, ,\\
\label{eq:K_2p}
	\mathcal{K}^{(2p+1)}&=\frac{m^{2}}{2a^{2p+1}}\langle \Phi\, (-\triangle)^{p+1/2} A_{0} +A_0\, (-\triangle)^{p+1/2} \Phi   \rangle\, . 
\end{align}
They can be expressed in terms of the corresponding mode functions as
\begin{align}
\label{eq:sp-Q2p}
	\mathcal{Q}^{(2p)}&=\int_{0}^{k_{h,L}}\frac{d k}{k}\frac{1}{2\pi^{2}I_{1}} \frac{k^{2p+3}}{a^{2p+2}}\,\big| \mathcal{D}_{L} \big| ^{2},\\
\label{eq:sp-F2p}
	\mathcal{F}^{(2p)}&=\int_{0}^{k_{h,L}}\frac{d k}{k}\frac{1}{2\pi^{2}I_{1}} \frac{k^{2p+3}}{a^{2p+2}}\,\big| \mathcal{A}_{L} \big| ^{2},\\
\label{eq:sp-K2p}
	\mathcal{K}^{(2p+1)}&=\int_{0}^{k_{h,L}}\frac{d k}{k}\frac{1}{2\pi^{2}I_{1}} \frac{k^{2p+4}}{a^{2p+3}}\, \operatorname{Re}\big[\mathcal{A}_{L}^{*}\,\mathcal{D}_{L} \big]\, .
\end{align}

Now, using either the definitions of functions $\mathcal{Q}^{(2p)}$, $\mathcal{F}^{(2p)}$, $\mathcal{K}^{(2p+1)}$ in coordinate space, Eqs.~\eqref{eq:Q_2p}--\eqref{eq:K_2p}, or the spectral representations above, we use Eqs.~\eqref{eq:eom-A0}--\eqref{eq:eom-Phi} or, equivalently, Eqs.~\eqref{eq:eom-AL-system}--\eqref{eq:eom-DL-system} in order to derive the equations of motion for the bilinear functions,
\begin{align}
\label{eq:dot_Q_2p}
	&\dot{\mathcal{Q}}^{(2p)} + 2H(p+3+M)\mathcal{Q}^{(2p)} + 2\mathcal{K}^{(2p+1)} =[\dot{\mathcal{Q}}^{(2p)}]_{\rm b}\,,\\
\label{eq:dot_F_2p}
	&\dot{\mathcal{F}}^{(2p)}+2H(p+1-M)\mathcal{F}^{(2p)}- 2\mathcal{K}^{(2p+1)}\nonumber\\
    &\hspace{3cm}-\frac{2m^{2}}{I_{1}}\mathcal{K}^{(2p-1)}=[\dot{\mathcal{F}}^{(2p)}]_{\rm b}\,,\\
\label{eq:dot_K_2p}
	&\dot{\mathcal{K}}^{(2p+1)}+2H(p+5/2) \mathcal{K}^{(2p+1)}+\mathcal{F}^{(2p+2)}\nonumber\\
    &\hspace{1.5cm}-\mathcal{Q}^{(2p+2)}-\frac{m^{2}}{I_{1}}\mathcal{Q}^{(2p)}=[\dot{\mathcal{K}}^{(2p+1)}]_{\rm b}\,.
\end{align}
Here again the terms on the right-hand side\,---\,the boundary terms\,---\,originate from the fact that only the modes with momenta $k\leq k_{h,L}$ are taken into account in the bilinear functions and that threshold momentum $k_{h,L}$ grows in time. In full analogy with the transverse case, the boundary terms can be derived using the spectral representation of bilinear functions in Eqs.~\eqref{eq:sp-Q2p}--\eqref{eq:sp-K2p} by differentiating them with respect to time and taking only the term coming from the upper integration boundary. The general expression follows from Eqs.~\eqref{X_integr}--\eqref{X_p_d} with the replacement $k_h\to k_{h,L}$. Finally, we get the following expressions:
\begin{align}
\label{eq:bound-Q}
	&[\dot{\mathcal{Q}}^{(2p)}]_{\rm b}=\frac{d \ln k_{h,L}}{d t}\,\frac{H^{2p+4}|R|^{2p+5}}{8\pi} \nonumber\\
    &\qquad\times\bigg|H^{(1)}_{\alpha+1}(|R|)-\frac{3/2+\alpha+M}{|R|} H^{(1)}_{\alpha}(|R|) \bigg|^{2}\, ,\\
\label{eq:bound-F}
	&[\dot{\mathcal{F}}^{(2p)}]_{\rm b}=\frac{d \ln k_{h,L}}{d t}\,\frac{H^{2p+4}|R|^{2p+5}}{8\pi}\Big|H^{(1)}_{\alpha}(|R|) \Big|^{2}\, ,\\
\label{eq:bound-K}
	&[\dot{\mathcal{K}}^{(2p+1)}]_{\rm b}=\frac{d \ln k_{h,L}}{dt}\,\frac{H^{2p+5}|R|^{2p+6}}{8\pi} \,\operatorname{Re}\bigg\{\!H^{(1)*}_{\alpha}(|R|) \nonumber\\
    &\qquad\times \bigg[H^{(1)}_{\alpha+1}(|R|)-\frac{3/2+\alpha+M}{|R|} H^{(1)}_{\alpha}(|R|)\bigg] \bigg\}\, .
\end{align}

The system of equations of the GEF is also infinite, as in the transverse case. Indeed, equations for $\mathcal{Q}^{(2p)}$ and $\mathcal{F}^{(2p)}$ contain $\mathcal{K}^{(2p+1)}$ while the equation for the latter contains $\mathcal{Q}^{(2p+2)}$ and $\mathcal{F}^{(2p+2)}$ etc. It can be truncated in two possible ways,
\begin{itemize}
    \item[(i)] at even order $2p_{\mathrm{max}}$ by imposing the relation
    \begin{equation}
        \mathcal{K}^{(2p_{\mathrm{max}}+1)} = \Big(\frac{k_{h,L}}{a}\Big)^2 \mathcal{K}^{(2p_{\mathrm{max}}-1)} 
    \end{equation}
    and, thus, closing equations for $\mathcal{Q}^{(2p_{\mathrm{max}})}$ and $\mathcal{F}^{(2p_{\mathrm{max}})}$;

    \item[(ii)] at odd order $(2p_{\mathrm{max}}+1)$ by imposing relations
    \begin{align}
        \mathcal{Q}^{(2p_{\mathrm{max}}+2)} = \Big(\frac{k_{h,L}}{a}\Big)^2 \mathcal{Q}^{(2p_{\mathrm{max}})}\, ,\nonumber\\
        \mathcal{F}^{(2p_{\mathrm{max}}+2)} = \Big(\frac{k_{h,L}}{a}\Big)^2 \mathcal{F}^{(2p_{\mathrm{max}})} \,
    \end{align}
    and closing equation for $\mathcal{K}^{(2p_{\mathrm{max}}+1)}$.
\end{itemize}
Analogously to the case of transverse modes, the justification follows from the structure of the spectra of quadratic gauge-field quantities with a large number of spatial derivatives; see the discussion around Eq.~\eqref{eq:truncation-transverse}.

Due to the presence of terms with lower order, i.e., the term $\mathcal{K}^{(2p-1)}$ in the equation for $\mathcal{F}^{(2p)}$ and the term $\mathcal{Q}^{(2p+2)}$ in the equation for $\mathcal{K}^{(2p+1)}$, the system of equations also extends to the negative orders. However, the fact that such terms are absent in the equation for $\mathcal{Q}^{(2p)}$ allows us to truncate the system already at $p=-1$. Therefore, the only bilinear quantities with negative orders are $\mathcal{Q}^{(-2)}$ and $\mathcal{K}^{(-1)}$. Moreover, we will see that they are not only the auxiliary quantities allowing us to close the GEF system of equations; one of them, $\mathcal{Q}^{(-2)}$ is related to the electric energy density and enters the background equations of motion.

%%%%%%%%%%%%%%%%%%%%%%%%%%%%%%%%%%%%%%%%%%%%%%%%%%%%%%%%%%%%%%%%%%%%%%%%%%

%%%%%%%%%%%%%%%%%%%%%%%%%%%%%%%%%%%%%%%%%%%%%%%%%%%%%%%%%%%%%%%%%%%%%%%%%%
\section{Full system of equations}
\label{sec:system}

In this section, we summarize all equations that we found in previous parts of the paper. Let us start from the background equations for the Hubble rate and scalar inflaton field. 
We need to express all gauge-field quantities entering these equations in terms of the transverse and longitudinal bilinear quantities. Since the transverse and longitudinal contributions never mix inside quadratic expressions, we can study them separately.

The quantities constructed from the magnetic field are the simplest ones since they contain only the transverse components,
\begin{align}
    I_1 \langle \boldsymbol{B}^2 \rangle =  I_1 \langle \boldsymbol{B}^2_\perp \rangle &= \mathcal{B}^{(0)}\, ,\\
    \frac{I_1}{2} \langle \boldsymbol{E}\cdot\boldsymbol{B} + \boldsymbol{B}\cdot\boldsymbol{E} \rangle &=  \frac{I_1}{2}\langle \boldsymbol{E}_\perp \cdot\boldsymbol{B}_\perp +\boldsymbol{B}_\perp \cdot\boldsymbol{E}_\perp \rangle \nonumber\\
    &= -\mathcal{G}^{(0)}\, .
\end{align}

The term with the electric field squared has to be split as $I_1 \langle \boldsymbol{E}^2 \rangle = I_1 \langle \boldsymbol{E}_\perp^2 \rangle + I_1 \langle \boldsymbol{E}_\parallel^2 \rangle$, where
\begin{align}
    I_1 \langle \boldsymbol{E}_\perp^2 \rangle &= \mathcal{E}^{(0)}\, ,\\
    I_1 \langle \boldsymbol{E}_\parallel^2 \rangle &= I_1 \langle [(-\triangle)^{-1/2}\operatorname{div}\boldsymbol{E}_\parallel]^2 \rangle =\nonumber\\
    &=\frac{m^4 a^2}{I_1}\langle A_0 (-\triangle)^{-1} A_0\rangle = \frac{m^2}{I_1} \mathcal{Q}^{(-2)}\, ,
\end{align}
and in the last line we used Eq.~\eqref{Proca_34} for the divergence of the electric field.

Further, we express the terms quadratic in the vector potential. The square of the 0 component simply reads
\begin{equation}
    m^2 \langle A_0^2\rangle = \mathcal{Q}^{(0)}\, .
\end{equation}
The square of the three-vector decomposes similarly to the electric field contribution,
\begin{align}
    \frac{m^2}{a^2} \langle \boldsymbol{A}_\perp^2 \rangle &= \frac{m^2}{I_1}\mathcal{B}^{(-2)}\, ,\\
    \frac{m^2}{a^2} \langle \boldsymbol{A}_\parallel^2 \rangle &= m^2\langle \Big[\frac{1}{a}(-\triangle)^{-1/2}\operatorname{div}\boldsymbol{A}_\parallel\Big]^2 \rangle =\nonumber\\
    &=m^2\langle \Phi^2\rangle = \mathcal{F}^{(0)}\, .
\end{align}

Collecting all necessary terms, we write the Friedmann equations~\eqref{eq:Friedmann-explicit}--\eqref{eq:Friedmann-2-explicit} in the form
\begin{align}
\label{eq:Friedmann-final}
	&3H^{2}M_{\mathrm{P}}^{2}=\rho_{\rm tot}=\frac{1}{2}\dot{\phi}^{2}+V(\phi) \nonumber\\
    &+ \frac{1}{2}\Big(\mathcal{E}^{(0)}\!+\mathcal{B}^{(0)}\!+\mathcal{Q}^{(0)}\!+\mathcal{F}^{(0)}\Big)+\frac{m^{2}}{2I_{1}}\Big(\mathcal{Q}^{(-2)}+\mathcal{B}^{(-2)}\Big) \, ,\\
\label{eq:Friedmann-2-final}
    &-(2\dot{H}+3H^{2})M_{\mathrm{P}}^{2}=p_{\rm tot}=\frac{1}{2}\dot{\phi}^{2}-V(\phi) \nonumber\\
    &+ \frac{1}{6}\Big(\mathcal{E}^{(0)}\!+\mathcal{B}^{(0)}\!+3\mathcal{Q}^{(0)}\!-\mathcal{F}^{(0)}\Big)+\frac{m^{2}}{6I_{1}}\Big(\mathcal{Q}^{(-2)}-\mathcal{B}^{(-2)}\Big).
\end{align}
The Klein--Gordon equation~\eqref{eq:KG} reads
\begin{align}
	\label{eq:KG2}
	&\ddot{\phi}+3H\dot{\phi}+\frac{d V}{d \phi}=\frac{1}{2I_{1}}\frac{dI_{1}}{d\phi}\Big(\mathcal{E}^{(0)}-\mathcal{B}^{(0)}+\frac{m^{2}}{I_{1}}\mathcal{Q}^{(-2)}\Big)\nonumber\\
    &-\frac{1}{I_{1}}\frac{dI_{2}}{d\phi}\mathcal{G}^{(0)}+\frac{1}{m}\frac{dm}{d\phi}\Big(\mathcal{Q}^{(0)}-\mathcal{F}^{(0)}-\frac{m^{2}}{I_{1}} \mathcal{B}^{(-2)}\Big)\, .
\end{align}
 
Finally, we also have to take the system of differential equations~\eqref{dot_E_n}--\eqref{dot_B_n} for transverse components of the gauge field, truncated at some finite order $n_{\rm max}$, and the system of equations~\eqref{eq:dot_Q_2p}--\eqref{eq:dot_K_2p} for the longitudinal modes, truncated at a finite order $2p_{\rm max}$ or $2p_{\rm max}+1$. 

In the next section, we apply our GEF developed in this and the previous sections to a few benchmark scenarios of inflationary vector-field production.

%%%%%%%%%%%%%%%%%%%%%%%%%%%%%%%%%%%%%%%%%%%%%%%%%%%%%%%%%%%%%%%%%%%%%%%%%%
\section{Numerical results}
\label{sec:results}

\subsection{Model specification}

We consider the simplest chaotic inflationary model with the quadratic inflaton potential
\begin{equation}
\label{eq:potential}
    V(\phi)=\frac{m_\phi^2 \phi^2}{2}\, ,
\end{equation}
with the mass parameter $m_\phi=6\times 10^{-6}\,M_{\rm P}$. Although this model is excluded by CMB observations~\cite{Planck:2018jri} due to its prediction of an excessively large tensor perturbation amplitude, we employ it here for illustrative purposes and for the sake of simplicity. Moreover, the parabolic form of the potential in Eq.~\eqref{eq:potential} provides a good approximation to many realistic inflationary potentials near their minima—that is, in the region where the inflaton approaches the end of inflation, which is particularly relevant for vector field production. 

Initial conditions for the inflaton field are set at roughly 61 $e$-folds from the end of inflation (without gauge fields) by setting $\phi(0)=\phi_0=15.5\,M_{\rm P}$; the corresponding value for the inflaton derivative is taken from the slow-roll attractor solution $\dot{\phi}(0)\approx-V'M_{\rm P}/\sqrt{3V}\big\vert_{\phi_0} = -\sqrt{2/3}m_{\phi}M_{\rm P}\approx -4.9\times 10^{-6}\,M_{\rm P}^2$.

In our numerical analysis, we consider the limit of a very light vector field, i.e.,
\begin{equation}
    \mu = \frac{m}{\sqrt{I_1}H}\ll 1\, .
\end{equation}
This allows us to explicitly omit all terms containing the $m^2/I_1$ factor in the Friedmann equations~\eqref{eq:Friedmann-final}--\eqref{eq:Friedmann-2-final}, Klein--Gordon equation~\eqref{eq:KG2} as well as in the GEF systems of equations~\eqref{dot_E_n}--\eqref{dot_B_n} and \eqref{eq:dot_Q_2p}--\eqref{eq:dot_K_2p}. Note that this does not correspond to switching to a massless vector field and discarding its longitudinal modes. We will see that if the mass is small but time dependent (in fact, inflaton dependent), there may be efficient vector-field production, especially of its longitudinal modes.  In other words, although we are setting $\mu\ll 1$, we keep $M=\dot{m}/(mH)$ finite.

In the low-mass approximation above, the transverse GEF system~\eqref{dot_E_n}--\eqref{dot_B_n} takes the form that is exactly the same as in the case of a massless gauge field. It was extensively studied in the literature previously, see Refs.~\cite{Sobol:2020lec,Gorbar:2021rlt,Durrer:2023rhc,Lysenko:2025pse,Domcke:2023tnn}. Note that effects of time-dependent mass do not enter the set of equations for transverse modes; they are only sensitive to the kinetic and axial couplings. On the other hand, the longitudinal modes of the vector field are totally insensitive to the axial coupling as it never enters the corresponding equations of motion. Moreover, the kinetic coupling only enters the boundary terms while also not appearing in the GEF system~\eqref{eq:dot_Q_2p}--\eqref{eq:dot_K_2p} explicitly. Since we are primarily interested in studying the production of longitudinal modes of comparable magnitude to the transverse ones, in our numerical analysis, we keep only the mass couplings to the inflaton and in some cases also the kinetic coupling. Thus, we always set $I_2(\phi)=0$ in the remaining part of the paper.

For the kinetic coupling, we assume the functional dependence proposed by Ratra~\cite{Ratra:1991bn} in the context of inflationary magnetogenesis,
\begin{equation}
\label{eq:Ratra-fn}
    I_1(\phi)=\exp\Big(\frac{\beta\phi}{M_{\rm P}}\Big)\, ,
\end{equation}
where $\beta$ is the dimensionless coupling constant of the model. In principle, in order to avoid the strong coupling problem, one should take only the values $\beta>0$. However, in our numerical analysis below, we will consider also some negative values just to see the qualitative differences in the picture of the vector-field production.%
\footnote{In a realistic setup that avoids the strong coupling problem, one can accommodate the coupling function with negative $\beta$ at finite time intervals if they are followed by the intervals with $\beta>0$ (e.g., in a case of a piecewise coupling function); see, e.g., Refs.~\cite{Ferreira:2013sqa,Sharma:2017eps,Sharma:2018kgs, Nandi:2021lpf,Cecchini:2023bqu} for the case of inflationary magnetogenesis.}%

The inflaton-dependent mass is also taken in the exponential form
\begin{equation}
\label{eq:mass-fn}
    m(\phi)=m_0\exp\Big(\frac{\beta\phi}{2M_{\rm P}}\Big)\, ,
\end{equation}
where $m_0\ll H$ and we use the same coupling constant $\beta$ as in the Ratra function~\eqref{eq:Ratra-fn} for the kinetic function (if the latter is also included; if not, this is just an independent coupling). This is primarily done just to minimize the number of independent parameters in our numerical setup. Also, as an extra motivation, for this choice of the kinetic and mass coupling functions, the inflaton is coupled in a universal way to the entire vector field Lagrangian, i.e.,
\begin{equation}
    \mathcal{L}_{\rm int}=I_{1}(\phi)\Big[-\frac{1}{4}F_{\mu\nu}F^{\mu\nu}+\frac{m_0^2}{2}A_{\mu}A^{\mu} \Big]\, .
\end{equation}
The exponential-type functions as in Eqs.~\eqref{eq:Ratra-fn}--\eqref{eq:mass-fn} can naturally arise in models where the vector field is nonminimally coupled to gravity. Following the seminal work~\cite{Turner:1987bw}, one can consider interactions of the form $R^p A_\mu A^\mu$ (as well as analogous couplings $R^q F_{\mu\nu}F^{\mu\nu}$ to the kinetic term). When such terms are embedded into $f(R)$ gravity or Higgs inflation and transformed to the Einstein frame, the scalaron or Higgs field plays the role of the inflaton and typically couples to the vector field through exponential functions. This mechanism has been explored, for example, in Refs.~\cite{Savchenko:2018pdr,Sobol:2021jag,Durrer:2022emo} for massless gauge fields, and its generalization to the massive case with a coupling $\sim R^p A_\mu A^\mu$ is straightforward.

Another, and arguably more natural, form of the mass coupling arises in models where the vector mass is generated through symmetry breaking, as in the Higgs mechanism; see, e.g., Refs.~\cite{Firouzjahi:2020whk,Emami:2009vd,Emami:2013bk}. In this case, the mass typically scales as $m(\phi)\propto \phi$. For such a coupling one finds $M = \frac{\dot\phi}{H\phi}=\frac{\sqrt{2\epsilon}}{\phi/M_{\rm P}}\ll 1$, since $\epsilon=-\dot{H}/H^2\ll 1$ during slow-roll inflation and $\phi/M_{\rm P}\gtrsim 1$ in the chaotic inflation model~\eqref{eq:potential} used in our numerical analysis. Therefore, gauge-field production is expected to be very similar to the case of a constant mass, i.e., $M=0$, which has already been studied in the literature (see, e.g., Ref.~\cite{Ema:2019yrd}). Although our method is fully capable of reproducing the gauge-field evolution in this scenario, no significant backreaction from the longitudinal modes is anticipated, and the result can be obtained straightforwardly using conventional mode-by-mode analyses in momentum space. For this reason, we do not include this type of coupling in our numerical study.

Having thus specified our model, we now switch to presenting the numerical results.

\begin{figure}[htb!]
\centering
\includegraphics[width=0.99\linewidth]{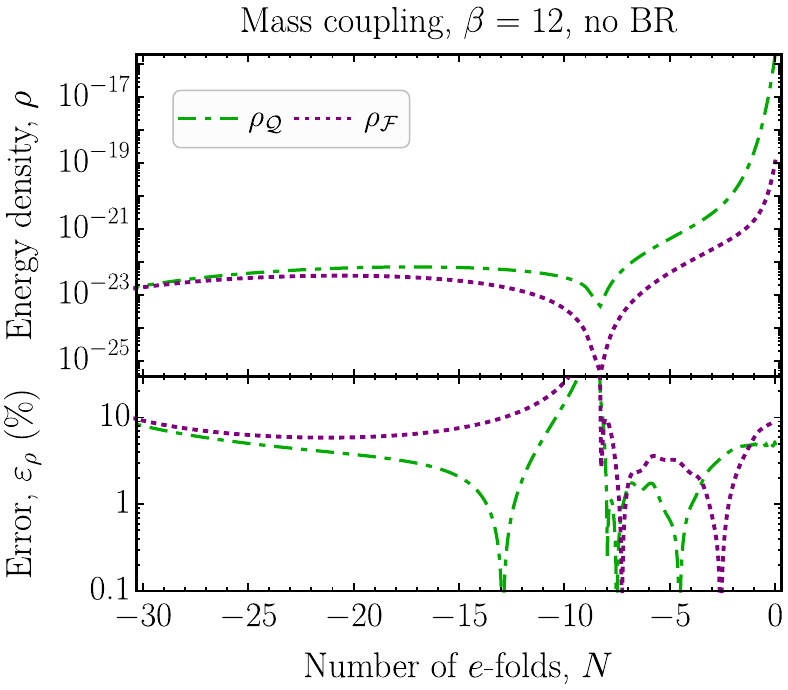}
\caption{Upper panel: the dependence of the energy densities $\rho_{\mathcal{Q}}$ (green dashed-dotted line) and $\rho_{\mathcal{F}}$ (purple dotted line) on the number of $e$-folds counted from the end of inflation in the model of mass coupling~\eqref{eq:mass-fn} of the vector field to inflaton with $\beta=12$, obtained by using the GEF truncated at the order $2p_{\rm max}+1=67$. Lower panel: evolution of the relative error of the GEF solution with respect to the mode-by-mode solution in momentum space.
\label{fig:mass-noBR}}
\end{figure}

\subsection{Pure mass coupling}

First, let us consider the case with only the mass coupling of the form~\eqref{eq:mass-fn} is present while the kinetic and axial couplings are trivial. In such a case, one can completely disregard the transverse modes and exclude their contributions from the background equations as they are not excited above the vacuum.

We start from the smaller coupling $\beta=12$ for which the longitudinal modes remain weak enough not to cause the backreaction on the inflationary background. The upper panel of Fig.~\ref{fig:mass-noBR} shows the evolution of the energy-density components
\begin{align}
\label{eq:rho-Q-def}
    \rho_{\mathcal{Q}}=\frac{m^2}{2}\langle A_0^2\rangle=\frac{1}{2}\mathcal{Q}^{(0)}\, ,\\
\label{eq:rho-F-def}
    \rho_{\mathcal{F}}=\frac{m^2}{2a^2}\langle \boldsymbol{A}_{\parallel}^2\rangle=\frac{1}{2}\mathcal{F}^{(0)}\, 
\end{align}
of the longitudinal modes (the green dashed--dotted and purple dotted lines, respectively) during the last 30 $e$-folds of inflation computed by using the GEF system of equations~\eqref{eq:dot_Q_2p}--\eqref{eq:dot_K_2p} truncated at the order $2p_{\rm max}+1=67$. The lower panel of this figure shows the relative error of the GEF solution compared to the numerical result obtained by treating separate Fourier modes in the momentum space and solving the system of Eqs.~\eqref{eq:eom-AL-system}--\eqref{eq:eom-DL-system} with boundary conditions in Eqs.~\eqref{eq:BC-AL}--\eqref{eq:BC-DL}. We will refer to the latter solution as \textit{the mode-by-mode (MbM)} solution and define the relative error as
\begin{equation}
\label{eq:error-def}
    \varepsilon_\rho = \bigg\vert\frac{\rho_{\rm GEF}}{\rho_{\rm MbM}}-1 \bigg\vert\times 100\,\%\, .
\end{equation}
Note that the relative error is typically of order a few percent and only takes somewhat larger values, or order 10\,\% in the vicinity of a sharp drop in the energy density curves shown in the upper panel of Fig.~\ref{fig:mass-noBR}. The latter drop occurs in a region where the horizon-crossing momentum $k_{h,L}$ stops growing because of the nonmonotonic behavior of the quantity $aH|R|$. Apart from the near-drop regions, the accuracy of the GEF method is very good and the method can be trustfully used for the studies of the vector DM production.

\begin{figure}[htb!]
\centering
\includegraphics[width=0.99\linewidth]{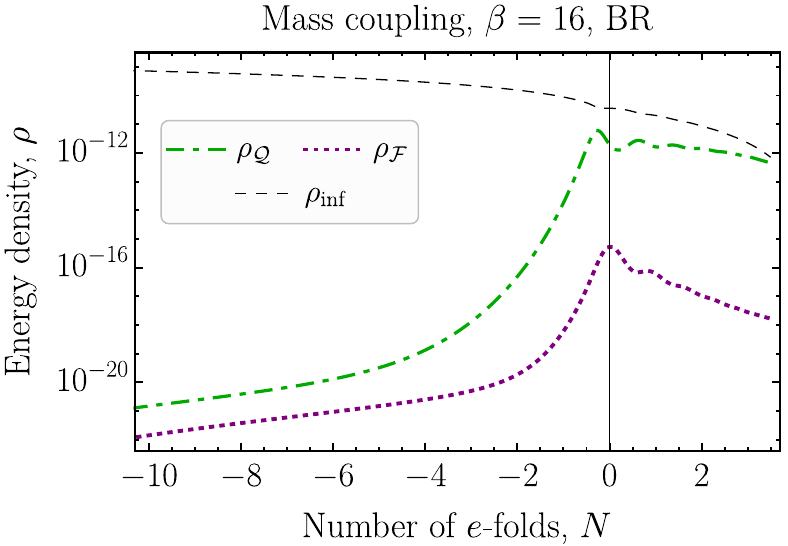}
\caption{Dependence of the energy densities $\rho_{\mathcal{Q}}$ (green dashed-dotted line) and $\rho_{\mathcal{F}}$ (purple dotted line) of the longitudinal modes of the vector field as well as the inflaton energy density $\rho_{\rm inf}$ (black dashed line) on the number of $e$-folds in the model of mass coupling~\eqref{eq:mass-fn} of the vector field to inflaton with $\beta=16$, obtained by using the GEF truncated at the order $2p_{\rm max}+1=67$. The point $N=0$ is chosen at the moment of time when inflation would have ended in the absence of gauge fields. The backreaction prolongs the accelerated expansion of the Universe by a few $e$-folds.
\label{fig:mass-withBR}}
\end{figure}

Next, in Fig.~\ref{fig:mass-withBR}, we show the numerical results for the case of a stronger coupling, $\beta=16$. Here we also plot the inflaton energy density $\rho_{\rm inf}=\dot{\phi}^2/2+V(\phi)$ by the thin black dashed line. The produced vector field appears to be strong enough to change the background inflationary evolution and extend the accelerated expansion of the Universe by a few $e$-folds. Indeed, the point $N=0$ on the horizontal axis is chosen at the moment of time when inflation would have ended in the absence of gauge fields. As one can see from the figure, inflation still lasts a few $e$-folds more after that if the presence of the vector field is taken into account. In the backreaction regime, the gauge-field energy density stops growing and reveals damped oscillatory behavior, which is very analogous to the backreaction regime for the massless gauge fields with the kinetic and axial couplings to the inflaton reported, e.g., in Refs.~\cite{Sobol:2020lec,Domcke:2020zez,Gorbar:2021rlt,Lysenko:2025pse}.

\begin{figure}[htb!]
\centering
\includegraphics[width=0.99\linewidth]{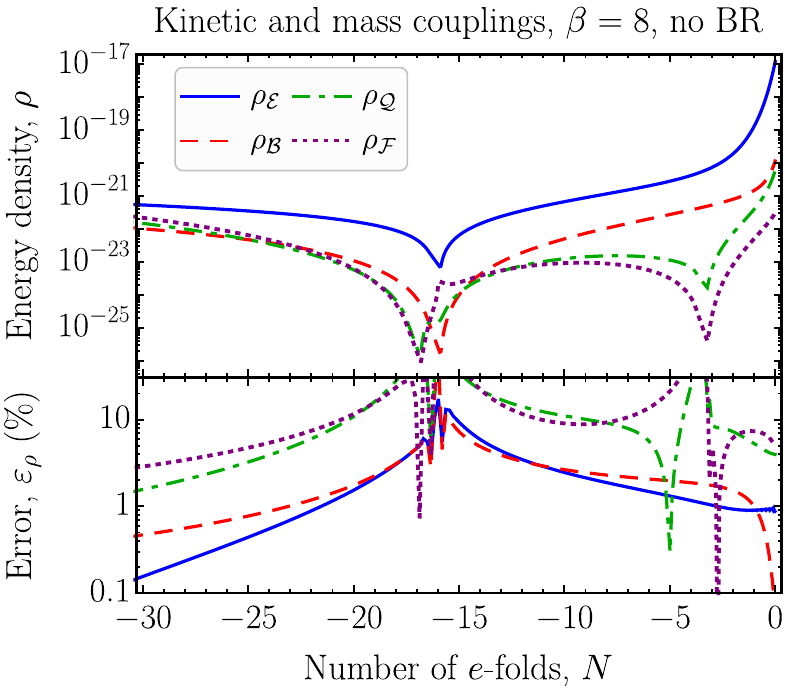}
\caption{Upper panel: the dependence of the energy densities $\rho_{\mathcal{E}}$ (blue solid line), $\rho_{\mathcal{B}}$ (red dashed line), $\rho_{\mathcal{Q}}$ (green dashed-dotted line), and $\rho_{\mathcal{F}}$ (purple dotted line) on the number of $e$-folds counted from the end of inflation in the model of kinetic~\eqref{eq:Ratra-fn} and mass couplings~\eqref{eq:mass-fn} of the vector field to inflaton with $\beta=8$, obtained by using the GEF truncated at the orders $n_{\rm max}=41$ for the transverse modes and  $2p_{\rm max}+1=41$ for the longitudinal modes. Lower panel: evolution of the relative error of the GEF solution with respect to the mode-by-mode solution in momentum space.
\label{fig:km-noBR-pos}}
\end{figure}

\begin{figure}[htb!]
\centering
\includegraphics[width=0.99\linewidth]{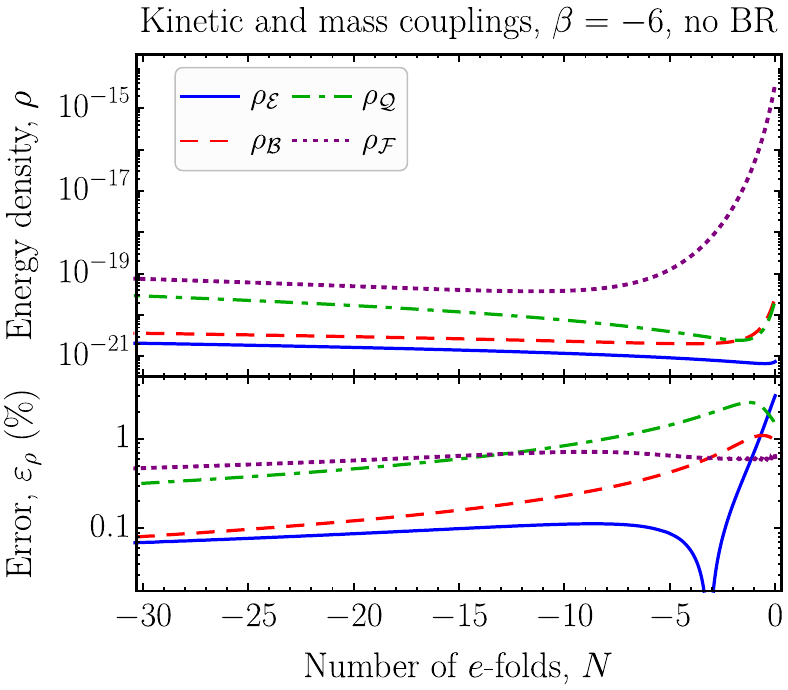}
\caption{Same quantities as shown in Fig.~\ref{fig:km-noBR-pos} for the case of a negative coupling constant $\beta=-6$. The GEF system was truncated at the orders $n_{\rm max}=31$ for the transverse modes and  $2p_{\rm max}+1=31$ for the longitudinal modes.
\label{fig:km-noBR-neg}}
\end{figure}

\subsection{Kinetic and mass couplings}

In the case when both the kinetic and mass couplings are present, also the transverse modes start being enhanced. We introduce two components of the energy density characterizing the contributions of the transverse modes of the vector field to the total energy density as follows:
\begin{align}
\label{eq:rho-E-def}
    \rho_{\mathcal{E}}=\frac{I_1}{2}\langle \boldsymbol{E}_{\perp}^2\rangle=\frac{1}{2}\mathcal{E}^{(0)}\, ,\\
\label{eq:rho-B-def}
    \rho_{\mathcal{B}}=\frac{I_1}{2}\langle \boldsymbol{B}^2\rangle=\frac{1}{2}\mathcal{B}^{(0)}\, .
\end{align}

Again, we start from the case of weak enough coupling so that the produced vector fields do not change the background inflationary evolution. Figure~\ref{fig:km-noBR-pos} shows the time evolution of the energy densities for the case of $\beta=8$ while the results for $\beta=-6$ are shown in Fig.~\ref{fig:km-noBR-neg}. Here the blue solid lines denote the transverse electric energy density $\rho_\mathcal{E}$; the red dashed lines show the evolution of the magnetic energy density $\rho_\mathcal{B}$; the green dashed-dotted and purple dotted lines, as earlier, denote the longitudinal components $\rho_\mathcal{Q}$ and $\rho_{\mathcal{F}}$, respectively. The lower panels of Figs.~\ref{fig:km-noBR-pos} and \ref{fig:km-noBR-neg} show the relative error of the GEF solutions with respect to the MbM treatment in the momentum space.

For the positive value of the coupling constant, $\beta=8$, the electric transverse modes clearly dominate among all gauge-field contributions. The longitudinal modes are subdominant with $\rho_\mathcal{Q}$ being slightly larger than $\rho_\mathcal{F}$ (the same hierarchy as in the pure mass coupling for $\beta=12$, see Fig.~\ref{fig:mass-noBR}). Here again, the error for the dominant energy density, $\rho_\mathcal{E}$, remains small, mostly of order 1\,\%, only reaching the values $\sim 10\,\%$ in a close vicinity of the sharp drop in the energy density. Its origin, similarly to the previous discussion in the case of pure mass coupling, is in the abrupt freezing of the $k_h$ evolution in the region when $aHr$ has a nonmonotonic behavior. Note that for longitudinal modes the sharp drop is also present (related to $k_{k,L}$ freezing in this case) and happens approximately 1 $e$-folding earlier. The subdominant contributions are determined by GEF with a slightly larger error, from a few to 10\,\%, and reaching up to 30\,\% error in the vicinity of the peaks. We would like to emphasize that this increase in the error is absolutely expected in this region as we are comparing two sharply peaked solutions (the ones from the GEF and MbM approaches). Indeed, any tiny mismatch between them leads to a large error as both functions sharply drop by two orders of magnitude near the minimum. In the case of monotonic functions, e.g., for a negative coupling constant $\beta=-6$ shown in Fig.~\ref{fig:km-noBR-neg}, the error of the GEF solution remains well below 10\,\% for all components of the energy density, including the subleading ones, while that for the dominant contribution always remains at the level of 1\,\%.

\begin{figure}[t]
\centering
\includegraphics[width=0.98\linewidth]{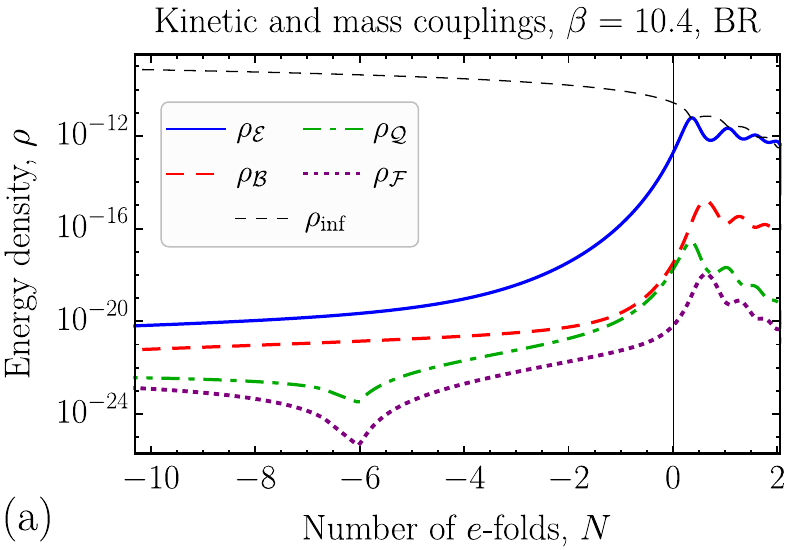}
\caption{The evolution of the different components of the energy density of the vector field: $\rho_{\mathcal{E}}$ (blue solid line), $\rho_{\mathcal{B}}$ (red dashed line), $\rho_{\mathcal{Q}}$ (green dashed-dotted line), $\rho_{\mathcal{F}}$ (purple dotted line), and the inflaton energy density $\rho_{\rm inf}$ (thin black dashed line) in the model of kinetic~\eqref{eq:Ratra-fn} and mass couplings~\eqref{eq:mass-fn} of the vector field to inflaton with $\beta=10.4$, obtained by using the GEF truncated at the orders $n_{\rm max}=63$ for the transverse modes and  $2p_{\rm max}+1=63$ for the longitudinal modes. The point $N=0$ is chosen at the moment of time when inflation would have ended in the absence of gauge fields.
\label{fig:km-BR-pos}}
\end{figure}
\begin{figure}[t]
\centering
\includegraphics[width=0.98\linewidth]{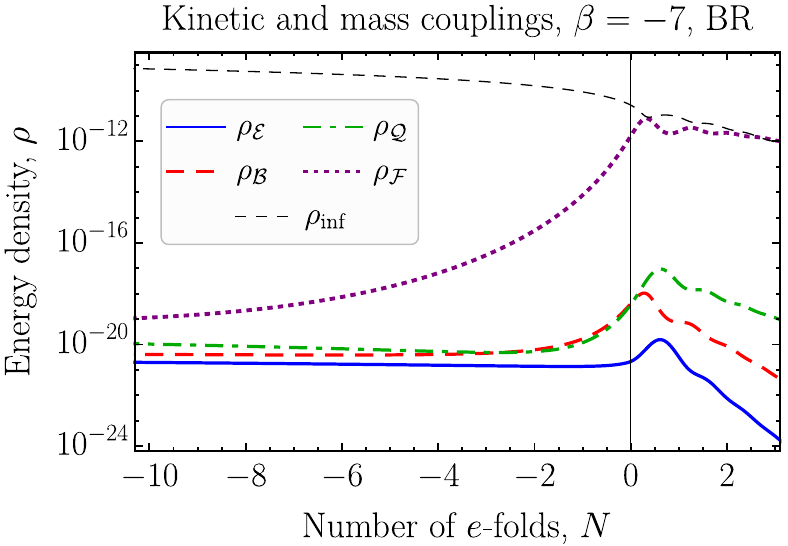}
\caption{Same quantities as shown in Fig.~\ref{fig:km-BR-pos} for the case of a negative coupling constant $\beta=-7$. The GEF system was truncated at the orders $n_{\rm max}=171$ for the transverse modes and  $2p_{\rm max}+1=171$ for the longitudinal modes.
\label{fig:km-BR-neg}}
\end{figure}

What is also interesting in the case of the negative coupling constant is that the hierarchy of the energy densities is fully flipped: from transverse to longitudinal modes, but also within each class. Now, the longitudinal component $\rho_{\mathcal{F}}$ is the largest one while the electric energy density $\rho_{\mathcal{E}}$ is the smallest. This flipping inside each class may be connected with the fact that both $\gamma$ and $M$ parameters now change their signs and an observation that $\gamma$ enters the GEF equations for $\mathcal{E}^{(n)}$ and $\mathcal{B}^{(n)}$ with different signs and $M$ enters the GEF equations for $\mathcal{Q}^{(2p)}$ and $\mathcal{F}^{(2p)}$ with different signs. Thus, flipping the sign of $\gamma$ and $M$ changes damping to enhancement and vice versa. 

Now, we switch to the cases with stronger couplings. The results for $\beta=10.4$ are shown in Fig.~\ref{fig:km-BR-pos} while those for $\beta=-7$ are shown in Fig.~\ref{fig:km-BR-neg}. Here, the hierarchy between different components of the energy density is the same as in the corresponding cases with weaker couplings. However, for stronger couplings, the produced fields are able to significantly impact the inflationary dynamics. Again, as in the pure mass coupling case, we see that the energy density of the vector field becomes comparable to that of the inflaton field. The time evolution of energy-density components reveals damped oscillations that probably result from the retardation between the changes in the inflaton velocity and the response of the gauge-field subsystem.

The accelerated inflationary expansion lasts longer in the presence of the gauge field. Despite the fact that, for the low-mass vector field, its energy density must redshift as radiation, the energy transfer from the inflaton through the mass and kinetic couplings maintains the gauge-field energy density almost constant and, thus, realizes a vacuumlike state composed from the inflaton and the gauge fields. This extends the duration of inflation by a few $e$-folds.

%%%%%%%%%%%%%%%%%%%%%%%%%%%%%%%%%%%%%%%%%%%%%%%%%%%%%%%%%%%%%%%%%%%%%%%%%%

%%%%%%%%%%%%%%%% Conclusions %%%%%%%%%%%%%%%%%%%%%%%%%%%%%%%%%%%%%%%%%%%%%
\section{Conclusions}
\label{sec:conclusions}

This work explored the possibility that dark photons are generated from vacuum fluctuations during inflation. Such production arises (i) because the presence of a mass term in the Lagrangian automatically breaks conformal invariance, leading to efficient generation of longitudinal vector modes, and (ii) through direct couplings to the inflaton field via kinetic, axial, and field-dependent mass terms.

In our analysis, we neglected kinetic mixing between the visible and dark sectors, focusing instead on the production of a single massive Abelian vector field interacting with the inflaton. We studied the dynamics of this dark vector field in complete isolation from the Standard Model sector.

We extended the gradient-expansion formalism (GEF)\,---\,previously developed for describing massless Abelian gauge field generation during inflation\,---\,to include the longitudinal polarization of a massive vector field. To this end, we derived a coupled system of equations of motion for bilinear functions of the vector field and examined the nonlinear dynamics of massive vector field production, including backreaction on the inflationary background. The GEF method is fully equivalent to the conventional mode-by-mode treatment in momentum space. The main difference is technical: Instead of evolving separate Fourier modes (which all become coupled in the nonlinear regime of evolution), the GEF operates directly in position space and evolves a set of bilinear gauge-field correlators that automatically include the contributions of all Fourier modes simultaneously. This makes the method considerably faster and computationally less expensive, especially when the gauge-field production is strong and many modes become coupled in a nonlinear way. The evolution of both transverse and longitudinal polarization modes was analyzed with the GEF and the result was cross-checked by means of the mode-by-mode approach in Fourier space.

We applied this formalism to a low-mass vector field whose kinetic and mass terms couple to the inflaton through a Ratra-type exponential function. Our study focused on the production of transverse and longitudinal polarizations in a benchmark quadratic inflationary model. We found that if only the mass coupling is present, the transverse polarizations are not enhanced while the longitudinal one can constitute a significant fraction of the energy density of the Universe and, for sufficiently strong couplings, even backreact to the inflationary background evolution. In the presence of both kinetic and mass couplings, much richer physics can be observed. In particular, when the coupling function decreases, transverse polarizations dominate the energy density. Conversely, for an increasing coupling function, the longitudinal component grows rapidly and drives the system into the strong backreaction regime.

In the cases without backreaction, we performed computations by employing two methods: by the GEF and by treating separate Fourier modes in momentum space. By comparing the results of two approaches, we estimated the relative error of the GEF result. It appears to be of order 1\,\% for the dominant component of the energy density and a few percent for subleading contributions if all energy densities evolve monotonically. If, on the other hand, sharp peaks or drops are present in the evolution of the energy-density components, the error in these regions may increase to $\sim 20-30\,\%$ for subleading components. Still, this result is fully acceptable for numerical study of the vector DM production during inflation.

The results presented here, obtained under the assumption of vanishing kinetic mixing with the electromagnetic field, provide a foundation for future work. In particular, our next study will address models with nonzero kinetic mixing, which acts as a portal between the visible and dark sectors and opens the possibility of experimentally detecting dark photons.

%%%%%%%%%%%%%%%%%%%%%%%%%%%%%%%%%%%%%%%%%%%%%%%%%%%%%%%%%%%%%%%%%%%%%%%%%

%%%%%%%%%%%%% Acknowledgements %%%%%%%%%%%%%%%%%%%%%%%%%%%%%%%%%%%%%%%%%%
\acknowledgments
S.~I.~V. is grateful to Professor Oleg Ruchayskiy for his support, useful discussions, and kind hospitality at the Niels Bohr Institute, University of Copenhagen, Denmark, where the final part of this work was done. S.~I.~V. is supported by the Erasmus+ Mobility Grant No. KA171-2023.
The work of A.~V.~L. was supported by the Swiss National Science Foundation through the Ukrainian--Swiss Joint research project ``Cosmic waltz of baryonic and ultralight dark matter: interaction and dynamical interplay'' (Grant No.
IZURZ2\_224972).
The work of O.\,S.\ has received funding through the SAFE\,---\,Supporting At-Risk Researchers with Fellowships in Europe project, which is funded by the European Union. Views and opinions expressed are, however, those of the authors only and do not necessarily reflect those of the European Union, the European Research Executive Agency (REA). Neither the European Union nor the granting authority can be held responsible for them.

%%%%%%%%%%%%%%%%%%%%%%%%%%%%%%%%%%%%%%%%%%%%%%%%%%%%%%%%%%%%%%%%%%%%%%%%%

%%%%%%%%%%%%% Data availability %%%%%%%%%%%%%%%%%%%%%%%%%%%%%%%%%%%%%%%%%%
\bigskip
\section*{Data availability}
The data that support the findings of this article are openly available~\cite{data:2026}.

% %%%%%%%%%%%%%%%% APPENDICES %%%%%%%%%%%%%%%%%%%%%%%%%%%%%%%%%%%%%%%%%%%%%
\appendix

\begin{widetext}

\section{Equation of motion for longitudinal modes}
\label{app:longitudinal}

In this appendix, we discuss in more details the equation of motion for the longitudinal polarization of the vector field.

Starting from Eqs.~\eqref{eq:eom-AL-system}--\eqref{eq:eom-DL-system}, we exclude the function $\mathcal{D}_L$ and get the following equation of motion for the mode function $\mathcal{A}_{L}(t,k)$:
\begin{align}
\label{eq:eom-AL-1}
    \ddot{\mathcal{A}}_{L}&+ \bigg[H-2\frac{\dot{m}}{m} + \frac{2k^2 I_1}{k^2 I_1+m^2 a^2} \bigg(H+\frac{\dot{m}}{m}- \frac{\dot{I}_1}{2I_1}\bigg)\bigg] \dot{\mathcal{A}}_{L} + \bigg[\frac{k^{2}}{a^{2}} +\frac{m^{2}}{I_{1}}  \nonumber \\
    &+2\frac{\dot{m}^2}{m^2}-H\frac{\dot{m}}{m}-\frac{\ddot{m}}{m}- \frac{\ddot{I_{1}}+H\dot{I_{1}}}{2I_{1}}+\frac{\dot{I}_{1}^{2}}{4I_{1}^{2}}-\frac{2 k^2 I_1}{k^2 I_1+m^2 a^2} \bigg(\frac{\dot{m}^2}{m^2}+H\frac{\dot{m}}{m} +H\frac{\dot{I}_1}{2I_1}- \frac{\dot{I}_1^2}{4 I_1^2} \bigg) \bigg]\mathcal{A}_{L}=0\,.
\end{align}
Using the dimensionless parameters in Eqs.~\eqref{eq:gamma-def}--\eqref{eq:mu-def} and \eqref{eq:M-def}, we rewrite Eq.~\eqref{eq:eom-AL-1} in a more compact form
\begin{align}
\label{eq:eom-AL-2}
    \ddot{\mathcal{A}}_{L}&+ \frac{(k/aH)^2 (3-2\gamma)+\mu^2(1-2M)}{(k/aH)^2 +\mu^2} H\dot{\mathcal{A}}_{L}+\bigg[(k/aH)^2 +\mu^2\nonumber\\
    &+ \gamma^2 - \gamma(3-\epsilon_H-\epsilon_\gamma)- M^2-M(3- \epsilon_H-\epsilon_M) +\frac{2 \mu^2}{(k/aH)^2 +\mu^2} \big(M^2+M+\gamma-\gamma^2 \big) \bigg]H^2\mathcal{A}_{L}=0\,.
\end{align}
Introducing a canonically normalized mode function $\psi_L$ via Eq.~\eqref{eq:canonical-mode-fn}, we obtain Eq.~\eqref{eq:eom-psi-L} with the coefficient
\begin{multline}
\label{eq:Q-func}
    Q(t,k)=\frac{k^{2}}{a^{2}}+\frac{m^{2}}{I_{1}}+\frac{a^{2}m^{2}}{k^{2}I_{1}+a^{2}m^{2}} \bigg[-\frac{\ddot{I_{1}}}{2I_{1}} -\left(\frac{\ddot{a}}{a}+\frac{\ddot{m}}{m} \right)\frac{k^{2}I_{1}}{a^{2}m^{2}}  \bigg] + \\ +\frac{a^{4}m^{4}}{\left(k^{2}I_{1}+a^{2}m^{2}\right)^{2}}\Bigg[\Big(\frac{\dot{I_{1}}}{2I_{1}} \Big)^{\!\!2}\left(\frac{4k^{2}I_{1}}{a^{2}m^{2}}+1\right) - \frac{H\dot{I_{1}}}{2I_{1}}\left(\frac{7k^{2}I_{1}}{a^{2}m^{2}}+1\right) - \frac{\dot{I_{1}}}{2I_{1}}\frac{\dot{m}}{m} \frac{6k^{2}I_{1}}{a^{2}m^{2}} + \\+ H^{2} \left(-\,\frac{k^{2}I_{1}}{a^{2}m^{2}}+2\right)\frac{k^{2}I_{1}}{a^{2}m^{2}} +3H\frac{\dot{m}}{m}\left(-\,\frac{k^{2}I_{1}}{a^{2}m^{2}}+1\right)\frac{k^{2}I_{1}}{a^{2}m^{2}} + 3\Big(\frac{\dot{m}}{m}\Big)^2\!\cdot\frac{k^{2}I_{1}}{a^{2}m^{2}}   \Bigg].
\end{multline}
For the trivial kinetic coupling, $I_1\equiv 1$, this expression coincides with the one reported in Refs.~\cite{Marriott-Best:2025sez,LaRosa:2025woi}. In the limit $\mu\ll 1$ and for $k\gtrsim aH$, the expression simplifies to the one in Eq.~\eqref{eq:Q-approx}. Far inside the horizon, the equation of motion for $\psi_L$ in conformal time $\eta$ takes the oscillatorlike form $\psi''_L(\eta,k) + a^2 Q \psi_L(\eta, k)=0$.
To obtain a positive-frequency solution to this equation, we impose the Bunch--Davies boundary condition \eqref{eq:Bunch-Davies-BC-L}. Using Eq.~\eqref{eq:canonical-mode-fn}, we translate it to the boundary condition \eqref{eq:BC-AL} for $\mathcal{A}_L$. Further, using Eq.~\eqref{eq:eom-AL-system}, we derive the  boundary condition \eqref{eq:BC-DL} for $\mathcal{D}_L$.

By changing the function $\psi_L(z)=x^{1/2}f(x)$ where $x=-z$, we rewrite Eq.~\eqref{eq:eom-psiL-2} in the form of the Bessel equation:
\begin{equation}
    \frac{d^2 f}{dx^2} + \frac{1}{x}\frac{df}{dx} + \Big(1-\frac{1/4+R^2}{x^2}\Big)f=0\, .
\end{equation}
Using the boundary condition \eqref{eq:Bunch-Davies-BC-L}, we find a unique solution to this equation, which is expressed in terms of the Hankel function of the first kind; see Eq.~\eqref{eq:psi-L-solution}. 
\end{widetext}

% %%%%%%%%%%%%%%%%%%%%%%%%%%%%%%%%%%%%%%%%%%%%%%%%%%%%%%%%%%%%%%%%%%%%%%%%%

%%%%%%%%%%%%%%% Bibliography %%%%%%%%%%%%%%%%%%%%%%%%%%%%%%%%%%%%%%%%%%%%

\end{document}